\documentclass[aps,prb,twocolumn,superscriptaddress]{revtex4}

\usepackage{graphicx,bm,amssymb,amsmath,xcolor,pifont,soul}

\usepackage[linktocpage=true,colorlinks=true,pdfborder={0 0 0},linkcolor=blue,citecolor=red,filecolor=yellow,urlcolor=blue,bookmarks,pdfauthor={},]{hyperref}
\usepackage{ulem}
\usepackage{chngcntr}

\begin{document}

\title{Superconducting properties of MoTe$_2$ from the \textit{ab initio} anisotropic Migdal-Eliashberg theory}
	
\author{Hari Paudyal}
\affiliation{Department of Physics, Applied Physics, and Astronomy, Binghamton University-SUNY, Binghamton, New York 13902, USA}
\author{Samuel Ponc\'{e}}
\affiliation{Department of Materials, University of Oxford, Parks Road, Oxford OX1 3PH, United Kingdom}
\affiliation{Theory and Simulation of Materials (THEOS), \'Ecole Polytechnique F\'ed\'erale de Lausanne, CH-1015 Lausanne, Switzerland}
\author{Feliciano Giustino}
\affiliation{Oden Institute for Computational Engineering and Sciences, The University of Texas at Austin, Austin, Texas 78712, USA}
\affiliation{Department of Physics, The University of Texas at Austin, Austin, Texas 78712, USA}
\author{Elena R. Margine}
\email{rmargine@binghamton.edu}
\affiliation{Department of Physics, Applied Physics, and Astronomy, Binghamton University-SUNY, Binghamton, New York 13902, USA}
\date{\today}
	
\begin{abstract}
Molybdenum ditelluride (MoTe$_2$) is attracting considerable interest since it is the archetypal type-II Weyl semimetal and a candidate for topological superconductivity. We investigate the superconducting phase diagram of two MoTe$_2$ polymorphs using the {\it ab initio} anisotropic  Migdal-Eliashberg theory, and we show that the superconducting dome originates from the synergistic contribution of the density of states at the Fermi level and the transverse acoustic Te modes in the 1T$^\prime$ phase. We find that the electron and hole pockets carry trivial $s$-wave order parameters of slightly different magnitude, reminiscent of a two-gap structure as suggested by recent experiments. We suggest that a possible route for enhancing the superconducting critical temperature, and realizing $s_{+-}$ pairing, in the T$_d$ phase is to exploit its non-trivial band topology via electron doping.
\end{abstract}
\maketitle

\section{Introduction}

Molybdenum ditelluride (MoTe$_2$) is a member of the transition metal dichalcogenide family that has recently attracted significant attention since it hosts a wealth of exotic phases and emergent phenomena. Similar to the sister compound tungsten ditelluride (WTe$_2$)~\cite{ALI,PAN,KANG}, MoTe$_2$ displays large non-saturating magnetoresistance~\cite{KEUM,LEE,THIRUPATHAIAH} and pressure-driven superconductivity~\cite{QI,TAKAHASHI,HEIKES,GUGUCHIA,DISSANAYAKE}, and it is predicted to be a type-II Weyl semimetal~\cite{SUN,SOLUYANOV,WANG,TAMAI,DENG}. Since MoTe$_2$ is a prime candidate for realizing topological superconductivity and Marjorana fermions~\cite{GUGUCHIA,LI,LUO,NAIDYUK,DISSANAYAKE,HOSUR}, it is important to understand the interplay between the superconducting pairing mechanism, the Fermi surface (FS) topology, and the structural phase transition under applied pressure.

\begin{figure}[!]
	\centering
	\includegraphics[width=0.99\linewidth]{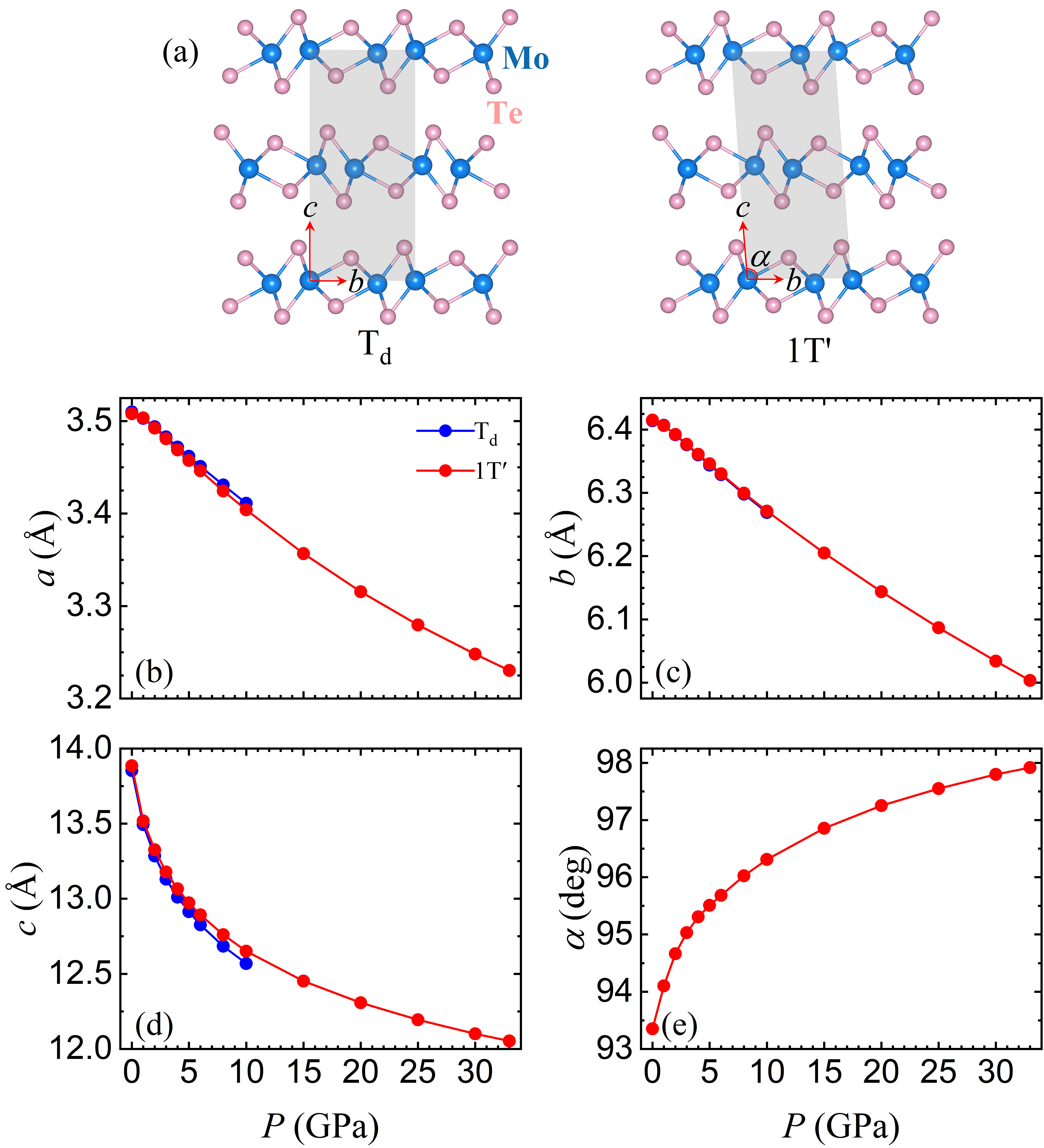}       
	\caption{\label{fig1} (a) Crystal structures of MoTe$_2$ in the T$_d$ and 1T$^\prime$ phase. (b)-(e) Pressure dependence of the structural parameters  for the T$_d$ (blue symbols) and 1T$^\prime$ (red symbols) phases.}
\end{figure}

The observed dome-like shape of the superconducting critical temperature ($T_{\rm c}$) with pressure and chemical doping has been associated with the structural transition from the non-centrosymmetric T$_d$ phase to the centrosymmetric 1T$^\prime$ phase~\cite{QI,TAKAHASHI,HEIKES,GUGUCHIA,CHEN,LEE,LI,MANDAL,CHO} shown in Fig.~\ref{fig1}(a), but several open questions remain: What is the nature of the superconducting gap in the two phases? Can the topological states take part in and enhance the superconducting paring in the T$_d$-phase? How does the critical temperature change when the Fermi level crosses a Weyl point?

Resistivity measurements have revealed that the $T_{\rm c}$ first increases sharply from 0.1~K at ambient pressure to approximately 5~K at pressures below 1~GPa, and then follows a gradual increase before reaching the maximum value of $T_{\rm c}=8.2$~K at 11.7~GPa~\cite{QI}. Since the T$_d$-to-1T$^\prime$ phase transition is suppressed at a critical pressure ($P_c$) between 1.5 and 4~GPa, this two-stage behavior was attributed to superconductivity in the T$_d$ and 1T$^\prime$ phases, respectively~\cite{QI}. In contrast, three subsequent transport studies have found a much lower critical pressure for the structural phase transition, and established the coexistence of the two phases at low temperature in the vicinity of $P_c$ as well as a strong superconductivity enhancement in the phase transition region~\cite{LEE,TAKAHASHI,HEIKES}. Takahashi {\it et al.} observed no evidence of the T$_d$ phase at 0.75~GPa down to 0~K~\cite{TAKAHASHI}, Heikes {\it et al.} reported  the suppression of the T$_d$ phase below 80~K at 0.82~GPa~\cite{HEIKES},  and Lee {\it et al.}  found that the structural transition temperature decreases to 58~K at 1.1~GPa, with no signature of the T$_d$ phase above 1.4~GPa~\cite{LEE}. The existence of a mixed phase region has been further supported by neutron-diffraction experiments, which have shown a $30\pm5\%$ volume fraction of the T$_d$ phase present at 1~GPa~\cite{HEIKES} and even evidence of the emergence of a new centrosymmetric T$_d^*$ phase across the T$_d$-1T$^\prime$ phase boundary~\cite{DISSANAYAKE}. 

Recent experiments have probed the nature of the superconducting gap in this system. Two isotropic $s$-wave superconducting gaps have been determined from muon-spin rotation experiments of MoTe$_2$ under pressure~\cite{GUGUCHIA} and specific heat measurements of S-doped MoTe$_{2-x}$S$_x$ ($x \sim 0.2$)~\cite{LI}. A two-band model has also been used to explain the superconducting properties in Se-substituted MoTe$_2$ thin films~\cite{LI2}. In order to make progress toward topological superconductivity, it is paramount to understand the origin of the superconducting dome in this compound, and to clarify the nature of the order parameter.

Here we analyze the nature of the superconducting dome and the pairing mechanism in MoTe$_2$ using the {\it ab initio} anisotropic Migdal-Eliashberg (ME) theory \cite{ALLEN1, MARGINE1}. We demonstrate that the ME theory reproduces quantitatively the measured superconducting dome, and we attribute the pressure-dependence of the critical temperature to the nonlinear variation of the transverse acoustic Te modes with compression. We show that the system exhibits $s$-wave pairing and an anisotropic gap structure that is reminiscent of a two-gap scenario, in line with recent experiments.

\section{Methods}

The \textit{ab initio} calculations were carried out with the Quantum ESPRESSO (QE)~\cite{QE} package. We employed relativistic norm-conserving pseudopotentials~\cite{NC-PP} with the Perdew-Burke-Ernzerhof (PBE)~\cite{PBE} exchange-correlation functional in the generalized gradient approximation, where the Mo $4d^5 5s^1$ and Te $5p^4 5s^2$ orbitals were included as valence electrons. To properly treat the long-range dispersive interactions, we used the non-local van der Waals (vdw) density functional optB86b-vdW~\cite{optB86b, vdW}. A plane wave kinetic-energy cutoff value of 60 Ry, a $\Gamma$-centered $12\times 6 \times 4$ Monkhorst-Pack \textbf{k}-mesh~\cite{k-mesh}, and a Methfessel and Paxton smearing~\cite{smearing} width of 0.02~Ry were used for the Brillouin-zone (BZ) integration. The atomic positions and lattice parameters were optimized until the self-consistent energy was converged within $2.7\times10^{-5}$ eV and the maximum Hellmann-Feynman force on each atom was less than 0.005~eV/\AA. For density of states and Fermi surface calculations, we used denser meshes of $24 \times 12 \times 8$ and $60 \times 40 \times 20$, respectively. The dynamical matrices and the linear variation of the self-consistent potential were calculated within density-functional perturbation theory~\cite{DFPT} on the irreducible set of a regular $4 \times 4 \times 3$ \textbf{q}-mesh.

The isotropic and anisotropic Migdal-Eliashberg (ME) formalism~\cite{ALLEN1,MARGINE1} was used to investigate the superconducting properties with the EPW code~\cite{Giustino2007,EPW}. To obtain the electron-phonon matrix elements on dense grids we use Wannier interpolation~\cite{WANN1,WANN2} on a uniform $\Gamma$-centered $8 \times 8 \times 3$ grid. Forty-four maximally localized Wannier functions (five $d$-orbitals for each Mo atom and three $p$-orbitals for each Te atom) were used to describe the electronic structure near the Fermi level ($E_{\rm F}$). Uniform $56\times 32 \times 16$ \textbf{k}-point and $28\times 16 \times 8$ \textbf{q}-point grids and an effective Coulomb potential $\mu^* = 0.1$ were used for solving the anisotropic ME equations. The Matsubara frequency cutoff was set to 0.2~eV and the Dirac deltas were replaced by Lorentzians of width 50~meV (electrons) and 0.1~meV (phonons).

\begin{figure*}[!]
	\centering
	\includegraphics[width=0.9\linewidth]{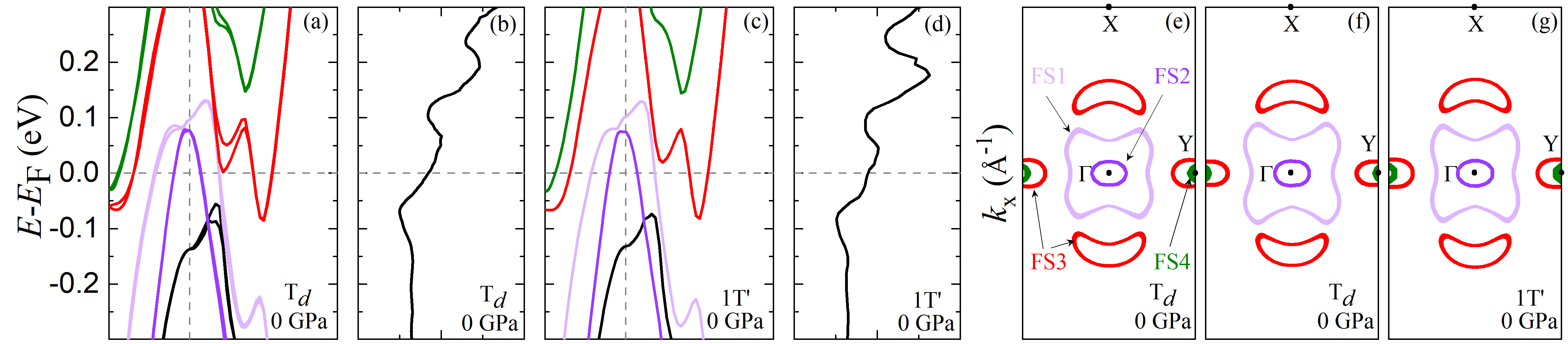}  
	\includegraphics[width=0.9\linewidth]{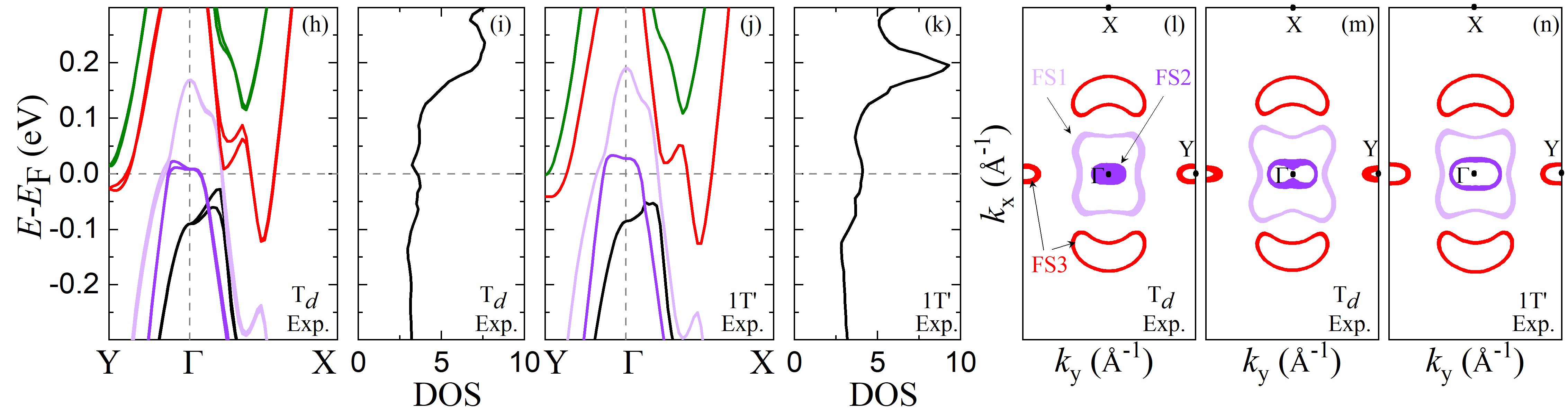}     
	\caption{\label{fig2} Top row: Calculated (a, d) electronic band structure, density of states (DOS)  in states/eV/u.c, and (e-g) cross section of the Fermi surfaces  in the $\Gamma$-$X$-$Y$ plane for the T$_d$  and 1T$^\prime$ phases at 0~GPa. A finite splitting of the bands can be observed in the T$_d$ phase due to the combined effect of SOC and lack of inversion symmetry (each pair of bands is shown with the same color in panel (a)).  The corresponding pair of  inner and outer FS sheets are shown in the (c, d) panels. The hole-like bands (light and dark purple curves) give rise to the $\Gamma$-centered  FS1 and FS2 hole pockets, while the electron-like bands (red and green curves) give rise to the FS3 and FS4 electron pockets. Bottom row: Same plots as in the top row but at the experimental unit cell parameters at ambient pressure.}
\end{figure*}

\section{Crystal Structure}

MoTe$_2$ crystallizes in three different phases: hexagonal-2H structure ($\alpha$-phase, space group P6$_3$/mmc, No.~194), monoclinic-1T$^\prime$ ($\beta$-phase, space group P2$_1$/m, No.~11), and orthorhombic-T$_d$ ($\gamma$-phase, space group Pmn2$_1$, No.~31). While both the semiconducting 2H and the semimetallic 1T$^\prime$ structures are stable at room temperature and ambient pressure, the semimetallic T$_d$ phase with broken inversion symmetry can be stabilized by cooling the 1T$^\prime$ phase down to approximately 250~K~\cite{QI}. In this work, we will focus on the semimetallic T$_d$ and 1T$^\prime$ phases. As shown in Fig.~\ref{fig1}(a), the two phases are closely related, both having a unit cell made of two Te-Mo-Te sandwiches and sharing the same in-plane crystal structure. A first-order phase transition between the T$_d$ and 1T$^\prime$ phase can be achieved by a slight relative sliding of the Te-Mo-Te layers. This structural transformation causes a small change in the vertical stacking, tilting the angle between the $a$ and $c$ lattice vectors in the 1T$^\prime$ unit cell.

At zero pressure, the calculated structural parameters are $a = 3.510$~\AA, $b = 6.414$~\AA, $c = 13.855$~\AA\, for T$_d$, and $a = 3.508$~\AA , $b = 6.415$~\AA, $c = 13.886$~\AA, $\alpha = 93.35 ^{\circ}$ for 1T$^\prime$, respectively. For comparison, the experimental lattice parameters at ambient pressure are $a = 3.477$~\AA, $b = 6.335$~\AA, $c = 13.889$~\AA\, for the T$_d$ phase~\cite{QI}, and $a = 3.469$~\AA , $b = 6.320$~\AA, $c = 13.860$~\AA, $\alpha = 93.917^{\circ}$ for the 1T$^\prime$ phase~\cite{BROWN}. Under compression, the in-plane lattice constants $a$ and $b$ decrease monotonically, while the out-of-plane lattice constant $c$ first drops sharply before beginning a slower descend around 5~GPa as shown in Fig.~\ref{fig1}(b-d). A gradual increase of the angle $\alpha$ with pressure can also be seen in Fig.~\ref{fig1}(e) for the 1T$^\prime$ phase. All these results are in good agreement with experimental and theoretical data reported in the literature~\cite{QI,HEIKES,GUGUCHIA}.

\section{RESULTS}
Figure~\ref{fig2}(a-g) shows the electronic structure of the T$_d$ and 1T$^\prime$ phases at 0~GPa with the inclusion of spin-orbit coupling (SOC). At 0~GPa, both structures display hole and electron bands crossing the Fermi level ($E_{\rm F}$) along the $Y$-$\Gamma$-$X$ direction. These bands give rise to a multi-sheet FS consisting of (i) a  butterfly-like hole pocket (FS1) enclosing a smaller ellipsoidal hole pocket (FS2) centered at $\Gamma$, and (ii) a two-dimensional (2D) electron pocket (FS3) surrounding a second 2D electron pocket (FS4) distributed along the $z$-axis. While in the 1T$^\prime$ phase the electron and hole pockets are doubly degenerate, in the T$_d$ phase the degeneracy is removed due to the broken inversion symmetry and each pocket is split in a pair of sheets as shown in Figs.~\ref{fig2}(a, h). This is a manifestation of the Rashba-Dresselhaus effect, as it can be seen near the band extrema at the $Y$ and $\Gamma$ points. This effect is most noticeable for the minor electron band above the Fermi level along the $\Gamma$-$X$ direction where a splitting of approximately 47~meV is observed (bands shown in red in Fig.~\ref{fig2}(a)). 

Results at higher pressures are shown in Supplemental Figs.~S1-S4~\cite{SM}. As pressure increases, the bands shift relative to $E_{\rm F}$ and the difference in the FS between the two phases becomes more pronounced. For instance at 4~GPa, the minor electron pocket along the $\Gamma$-$X$ direction shrinks in T$_d$, while it spreads out and touches the outer hole pocket in 1T$^\prime$.  Later we show how these changes in the electronic structure with applied pressure play an important role in the superconducting properties of the two phases.

\begin{figure*}[!]
	\centering
	\includegraphics[width=0.99\linewidth]{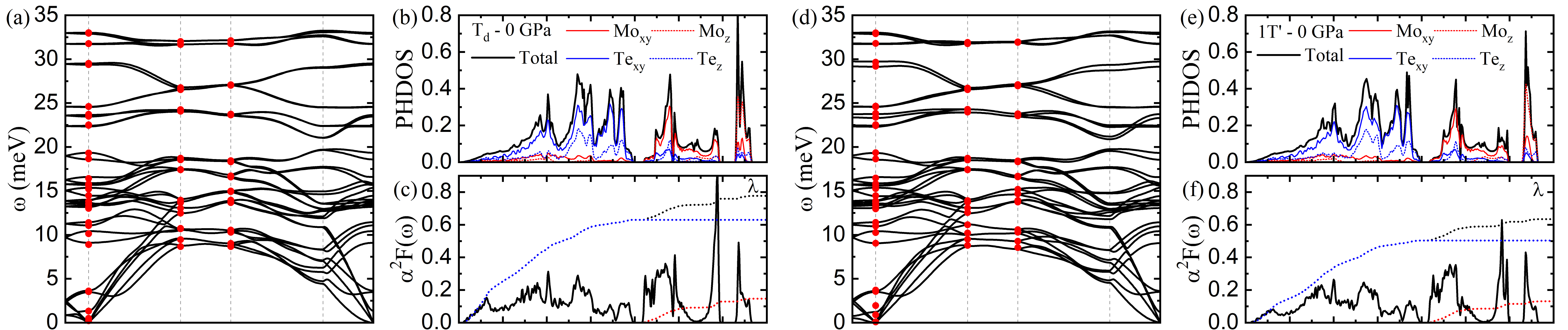}
	\includegraphics[width=0.99\linewidth]{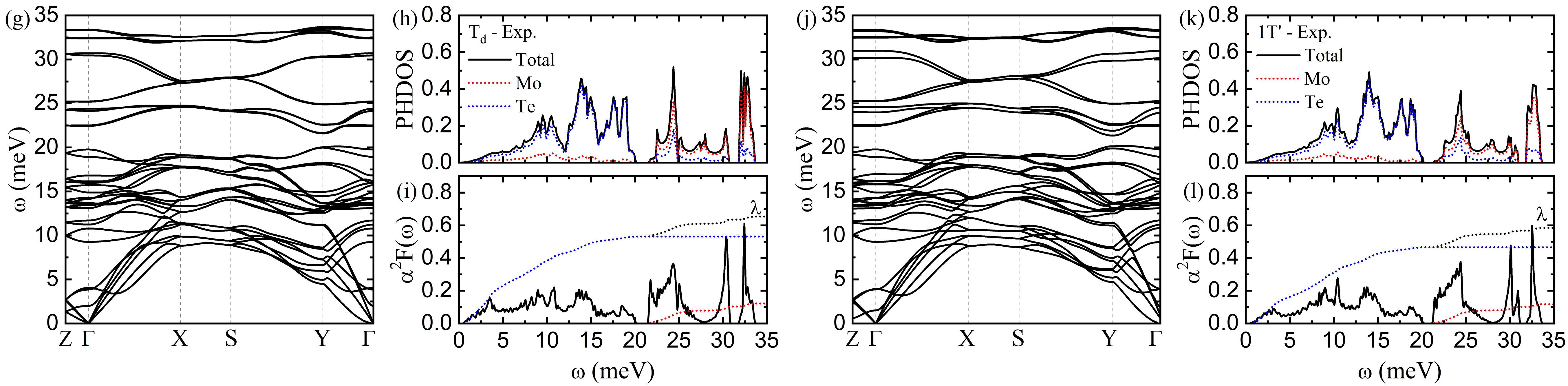}  
	\caption{\label{fig3} Top  row: Calculated (a, d) phonon dispersion, (b, e) phonon density of states (PHDOS), (c, f) isotropic Eliashberg spectral function $\alpha^2F(\omega)$, electron-phonon coupling strength $\lambda(\omega)$ for the T$_d$ and 1T$^\prime$ phases at 0 GPa. The total PHDOS (black line) is decomposed with respect to the in-plane and out-of-plane vibrations of the Mo and Te atoms. The total electron-phonon coupling strength (dotted black line) is decomposed with respect to the vibrations of the Mo (red lines) and Te (blue lines) atoms. The solid symbols represent the phonon frequencies calculated with spin-orbit coupling. Bottom row: Same plots as in the top row but at the experimental unit cell parameters at ambient pressure.}
\end{figure*}

The vibrational spectra and the phonon density of states (PHDOS) in the T$_d$ and 1T$^\prime$ phases at 0 GPa are shown in Figs.~\ref{fig3}(a, b, d, e).  The results at 0~GPa are in good agreement with experimental and theoretical results available in the literature~\cite{HEIKES,ZHANG,SONG,OLIVER,CHEN2} (see Supplemental Table~S1~\cite{SM}). To asses the importance of the SOC, we also calculated the phonon frequencies at specific high-symmetry points and, similar to a previous report~\cite{HEIKES}, we found the effect to be negligible (see solid red symbols in Figs.~\ref{fig3}(a, d). The two phases exhibit very similar phonon dispersions, and three energy regions can be distinguished with predominant contributions stemming from the in-plane Te$_{xy}$ vibrations (below 20~meV), the in-plane Mo$_{xy}$ vibrations (20-30~meV), and the out-of-plane Mo$_z$ vibrations (above 30~meV) as shown in Fig.~\ref{fig3}(b, e). 

\begin{figure}[!]
	\centering
	\includegraphics[width=0.99\linewidth]{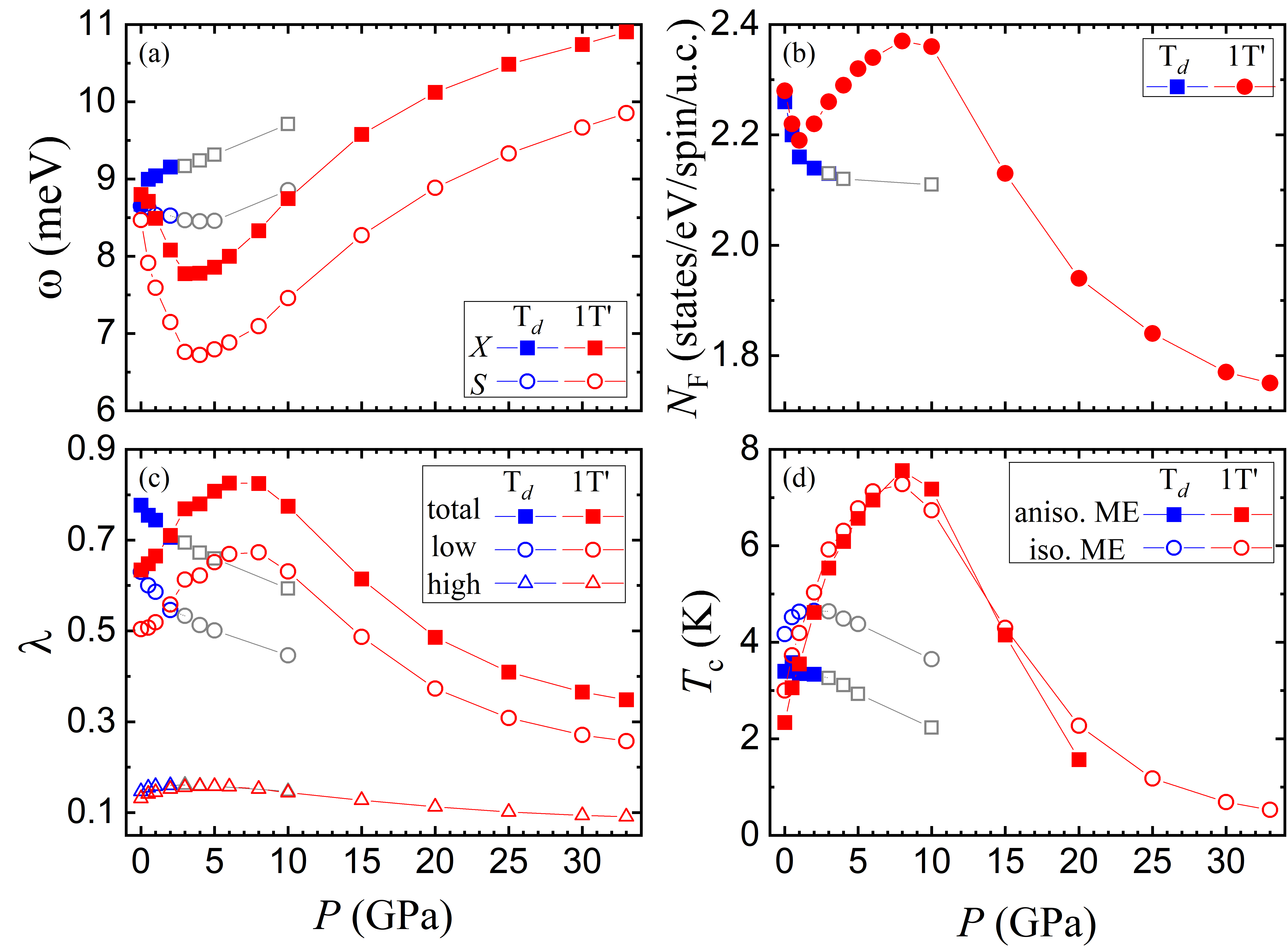} 
	\caption{\label{fig4} Variation with pressure of the (a) low-energy TA mode at $X$ and $S$, (b) $N_{\rm F}$,  (c) $\lambda$, and (d) $T_{\rm c}$ for the T$_d$ (blue) and 1T$^\prime$ (red) phases, respectively. In (c), squares represent the total  $\lambda$, and circles and triangles represent the contribution of the low- and high-energy modes.  In (d), squares and circles represent the $T_{\rm c}$ obtained from the numerical solutions of the isotropic and anisotropic ME equations. The data points in the pressure region where the T$_d$ phase is experimentally known to be suppressed are shown in gray.}
\end{figure}

With increasing pressure, all phonon modes harden across the whole Brillouin zone (BZ) (see Supplemental Figs.~S5-S6~\cite{SM}) with the exception of the lowest energy branch along the $X$-$S$ direction which softens at low pressures in the 1T$^\prime$ phase (see Fig.~\ref{fig4}(a)).  This transverse acoustic (TA) phonon mode is mainly characterized by displacements of Te atoms in both phases as illustrated in Supplemental Fig.~S7~\cite{SM}. A similar softening of the low-energy TA branch has been recently uncovered not only in the 1T$^\prime$ phase of MoTe$_2$~\cite{SI} but also in that of the sister compound WTe$_2$~\cite{LU}.

In order to clarify the nature of the electron-phonon coupling (EPC) in MoTe$_2$, we begin by examining the isotropic Eliashberg spectral function $\alpha^2F(\omega)$ and the cumulative EPC strength $\lambda(\omega)$. The results at 0~GPa are plotted in Figs.~\ref{fig3}(c, f), while the ones at higher pressures can be found in Supplemental Figs.~S5-S6~\cite{SM}. A comparison of the $\alpha^2F(\omega)$ with the PHDOS indicates that there is an increased coupling to the Te vibrational modes below 20~meV in both phases. The breakdown of the EPC strength into contributions arising from the low- and high-energy phonons demonstrates that at 0~GPa almost 80\%  of the total coupling comes from the Te modes. Fig.~\ref{fig4}(c) summarizes the dependence of $\lambda$ as a function of pressure. In the T$_d$ phase, $\lambda$ decreases with the applied pressure as the phonon modes harden and the density of states at the Fermi level ($N_{\rm F}$) decreases. In contrast, in the 1T$^\prime$ phase, $\lambda$ displays a dome-shaped variation and peaks between 6 to 8~GPa. This dependence correlates closely with the behavior of $N_{\rm F}$ and the low-energy modes in the 6-9~meV range along the $X$-$S$ direction. As shown in Figs.~\ref{fig4}(a)-(b) and also reported previously~\cite{SI}, the two exhibit an almost mirrored dependence over the full pressure range, which leads to their cooperative effect to $\lambda$. The low-energy TA mode was also found to have a significant contribution to $\lambda$ in the 1T$^\prime$ phase of WTe$_2$~\cite{LU}. In contrast to our results and those from Refs.~[\onlinecite{SI, LU}],  Heikes et al.~[\onlinecite{HEIKES}] observed no phonon softening in the 1T$^\prime$ phase of MoTe$_2$ up to 10~GPa. As a result, the estimated $\lambda$ was found to have a similar strength and to decrease slightly with pressure in both phases, giving an opposite trend for the $T_{\rm c}$ versus pressure behavior as compared to the one observed experimentally~\cite{HEIKES}.

\begin{figure*}[t]
	\centering
	\includegraphics[width=0.99\linewidth]{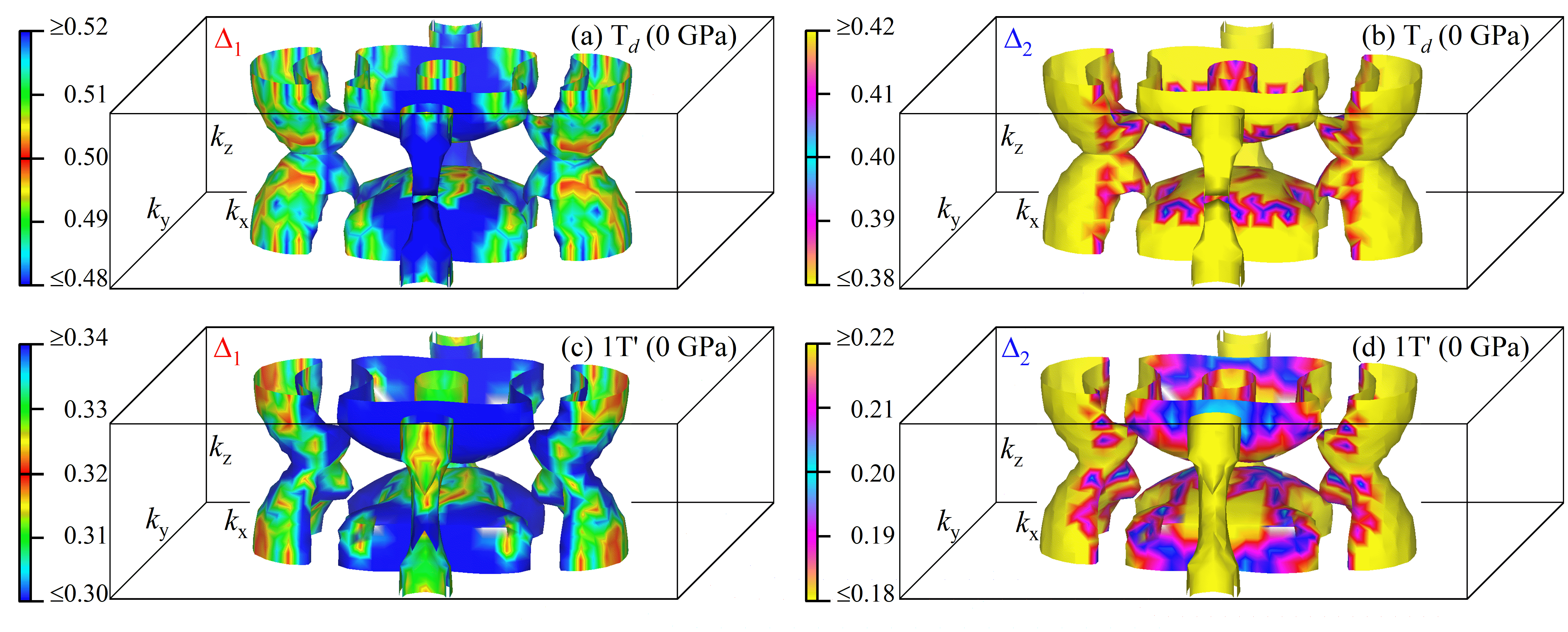}
	\includegraphics[width=0.99\linewidth]{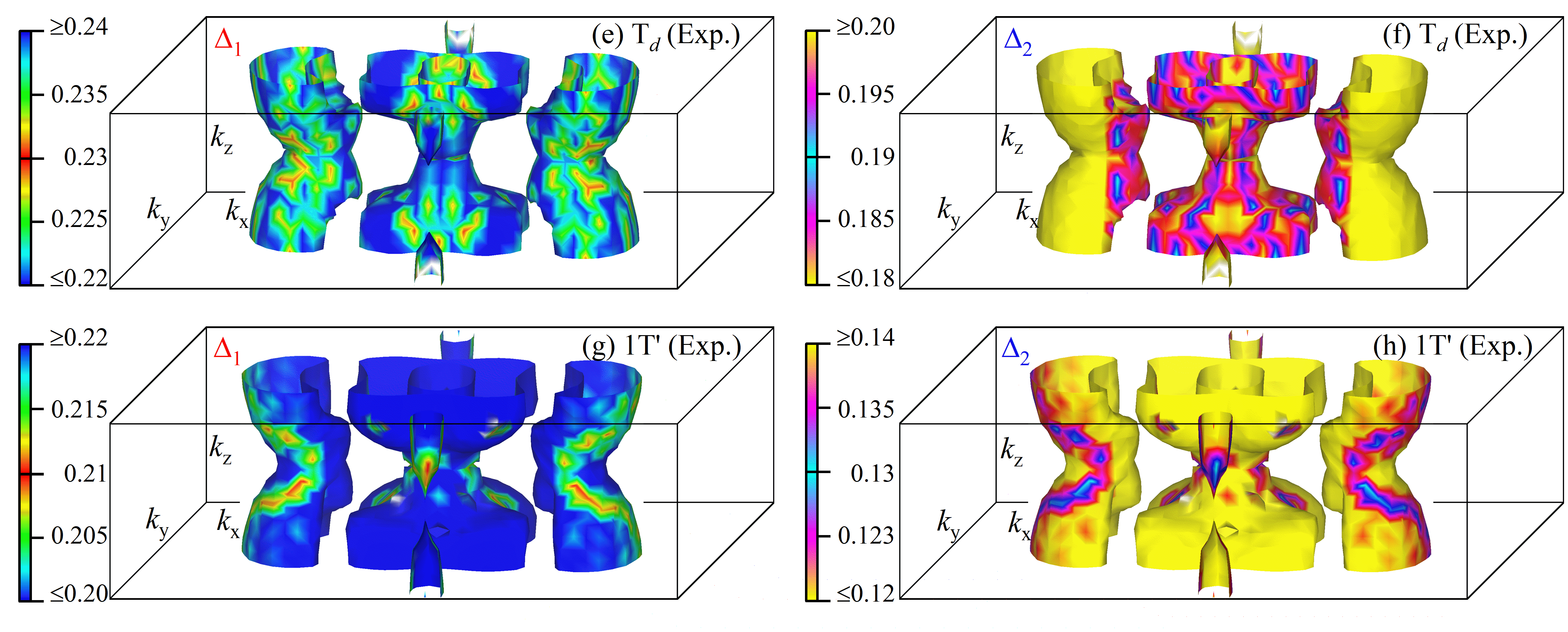}	
	\caption{\label{fig5} Momentum-resolved superconducting gap $\Delta_\textbf{k}$ (in meV) on the Fermi surface~\cite{VESTA} at 0.5~K for the T$_d$ and 1T$^\prime$ phases (a-d) at 0 GPa and (e-h) at the experimental lattice constants at ambient pressure. The distribution of the electronic states on the Fermi surface contributing the most to the $\Delta_1$ and $\Delta_2$ gaps is color-coded with Gaussians peaked at (a) 0.50~meV, (b) 0.40~meV, (c)  0.32~meV, and (d)  0.20~meV (b) 0.23~meV, (c) 0.19~meV, (e)  0.21~meV, and (f)  0.13~meV. The ranges of the color maps correspond to the $\Delta_1$ and $\Delta_2$ gaps indicated in Fig.~\ref{fig6}(a) and \ref{fig6}(d) at 0 GPa and Fig.~\ref{fig6}(c) and \ref{fig6}(f) at the experimental lattice constants at ambient pressure for the T$_d$ and 1T$^\prime$ phases.}
\end{figure*}

\begin{figure}[!]
	\centering
	\includegraphics[width=0.99\linewidth]{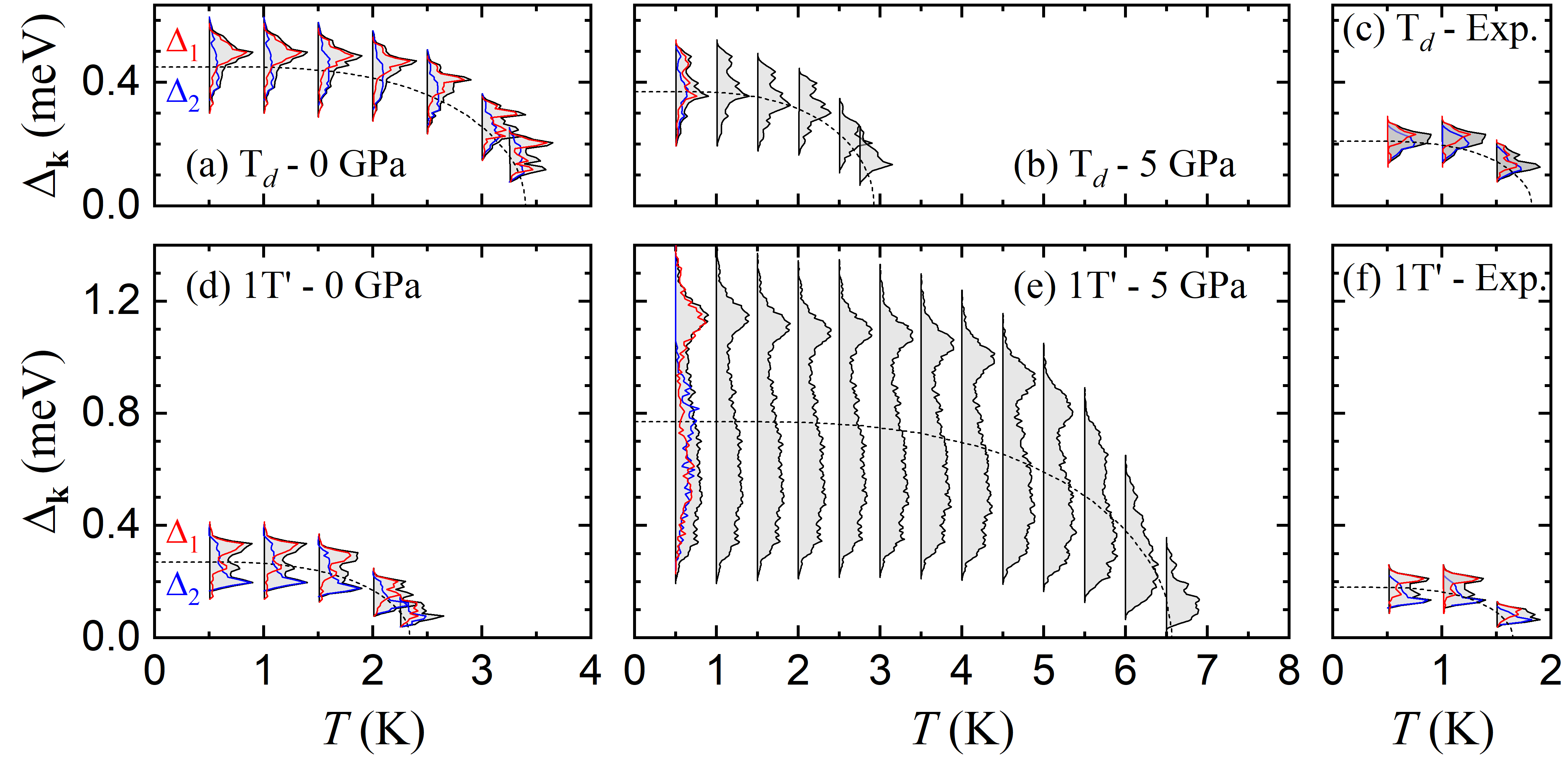}   
	\caption{\label{fig6} Energy distribution of the superconducting gap $\Delta_{\bf k}$ as a function of temperature for the T$_d$ (top panels) and 1T$^\prime$ (bottom panels) phases at 0~GPa, 5~GPa, and experimental lattice parameters at ambient pressure. The red and blue curves represent contributions to the $\Delta_1$ and $\Delta_2$ superconducting gaps associated with the electron and hole FS sheets, respectively. The dashed lines are fits obtained by solving numerically the BCS gap equation~\cite{BCS} using the average $\Delta_0$  and $T_{\rm c}$ from our first-principles calculations.}	
\end{figure}

To gain more insight into the superconductivity of MoTe$_2$, we compute the superconducting gap function $\Delta_{\bf k}$ on the FS by solving the anisotropic ME equations~\cite{ALLEN1,MARGINE1,EPW}. As shown in Fig.~\ref{fig5},  we find that a continuous anisotropic gap develops on the FS. Although we do not find definitive evidence of a two-gap structure, we observe that the distribution of the superconducting gap on the electron and hole Fermi pockets peak at slightly different energies (see rectangle in Supplemental Figs.~S3-S4~\cite{SM}). In particular,  while the superconducting gap $\Delta_1$ associated  with the electron sheets FS3 and FS4 spreads over the full range of the energy distribution of  the $\Delta_2$ gap associated with the hole sheets FS1 and FS2 (red and blue lines in Figs.~\ref{fig6}(a, d),  we also see that $\Delta_1$ is mostly concentrated in the upper half of the energy gap distribution and has a pronounced peak near the maximum of $\Delta_{\bf k}$.

\begin{figure}[t]
	\centering
	\includegraphics[width=\linewidth]{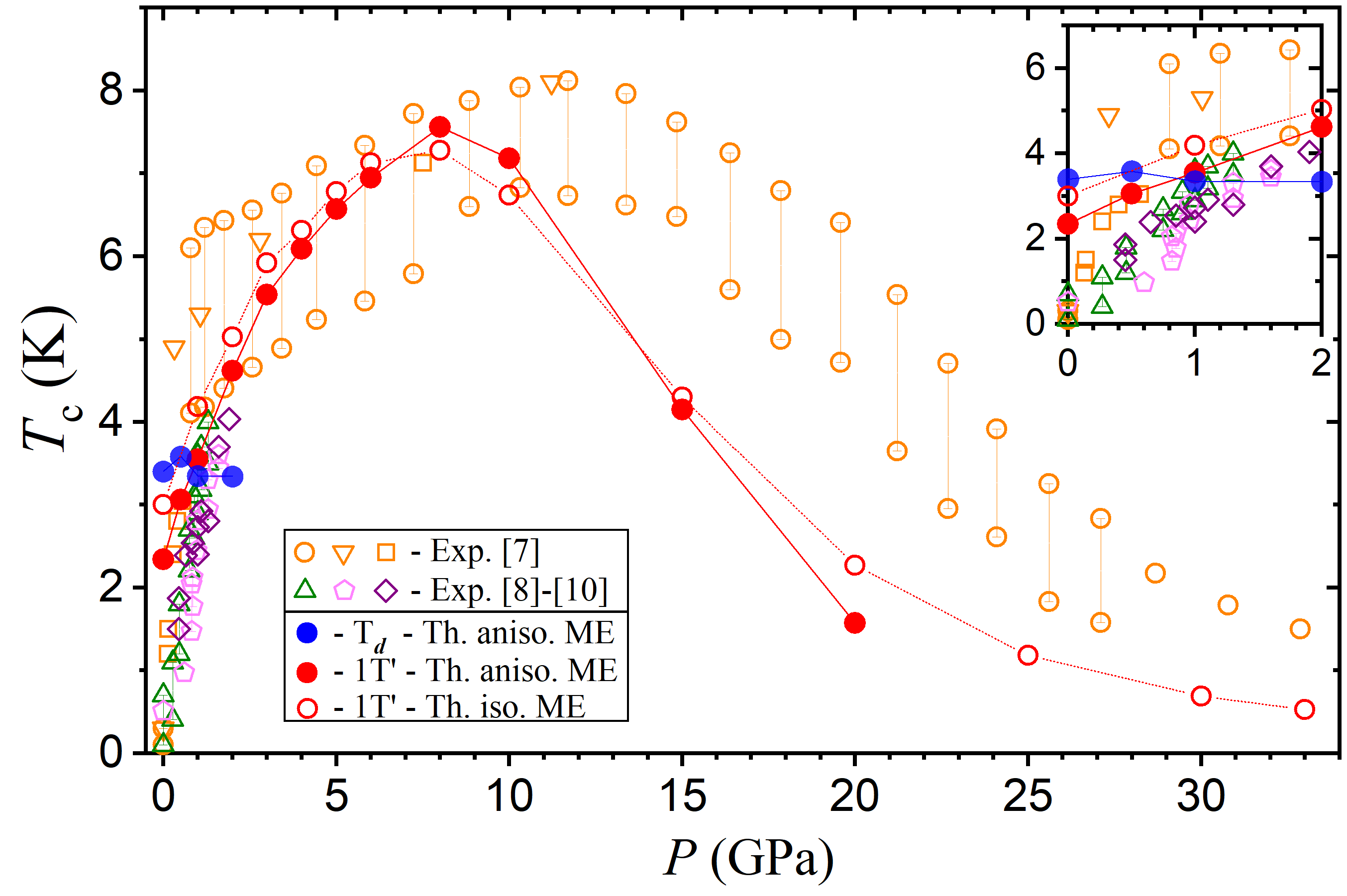} 
	\caption{\label{fig7} Comparison between experimental and theoretical results for $T_{\rm c}$  as a function of pressure.  The experimental data are from Ref.~[\onlinecite{QI}] (orange circles and triangles from electrical resistivity  and orange squares from magnetization measurements),  Ref.~[\onlinecite{TAKAHASHI}] (green triangles from electrical resistivity measurements),  Ref~[\onlinecite{HEIKES}] (magenta pentagons from neutron scattering measurements), and Ref.~[\onlinecite{GUGUCHIA}] (purple diamonds from AC-susceptibility measurements). The blue and red symbols represent the theoretical data for the T$_d$ and 1T$^\prime$ phases. For 1T$^\prime$, the filled and open red circles represent the solutions of the anisotropic and isotropic ME equations (above 20~GPa the gap is too small to be resolved at the anisotropic level). The vertical lines are a guide to the eye for differences in onset and zero-resistivity $T_{\rm c}$ in experimental data at various pressures.}
\end{figure} 

As the pressure is increased to 5~GPa, the anisotropic structure of the superconducting gap in the T$_d$ phase is only slightly affected,  both $\Delta_1$ and $\Delta_2$ gaps continuing to overlap over the full bandwidth of $\Delta_{\bf k}$ (Fig.~\ref{fig6}(b)). In contrast,  the energy distribution of the superconducting gap in the 1T$^\prime$ phase undergoes substantial changes as the hole and electron pockets spread out more and start to merge (Fig.~\ref{fig6}(e)).  The anisotropy in the 1T$^\prime$ phase becomes so pronounced that at 5~GPa the energy distribution of the gap is four times larger than that at 0~GPa. Above 10~GPa the spread in $\Delta_{\bf k}$ decreases, at 20~GPa the bandwidth being again comparable to the one found at zero pressure (see Supplemental Figs.~S8-S9~\cite{SM}). Another noticeable pressure-dependent characteristic is the narrowing of the energy profile of $\Delta_2$ with respect to $\Delta_1$ and its shift toward the lower energy region of $\Delta_{\bf k}$.

\begin{figure*}[t]
	\centering
	\includegraphics[width=0.99\linewidth]{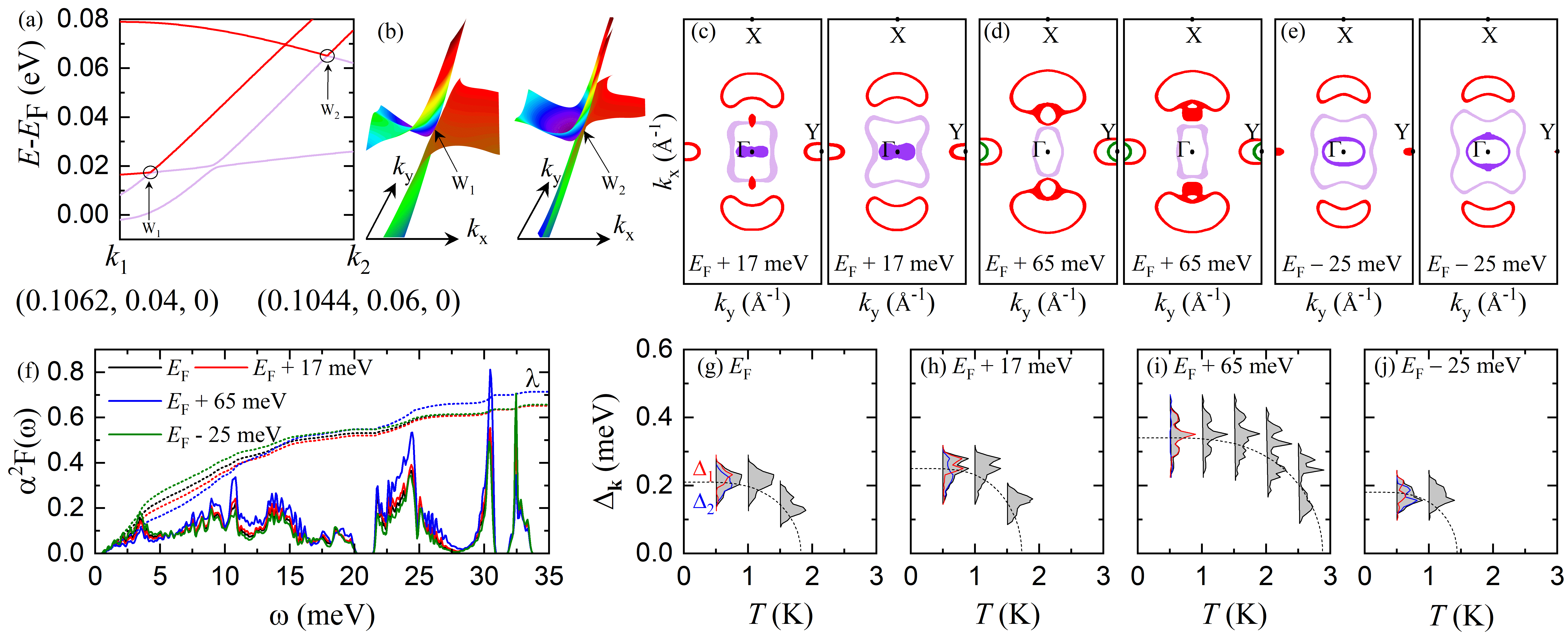}   
	\caption{\label{fig8} (a-b) Band structure and energy isosurfaces revealing the Weyl points  W$_1$ and W$_2$ at the intersections of electron and hole pockets.  W$_1$ (chirality +1) and W$_2$ (chirality -1) lie 17~meV and  65~meV above the Fermi level. The coordinates of the Weyl points are W$_1$ = (0.10599, 0.01028, 0) $\frac{2\pi}{a}$ and W$_2$ = (0.10463, 0.05305, 0)$\frac{2\pi}{a}$.  Cross section of the Fermi surfaces in the $\Gamma$-$X$-$Y$ plane calculated by shifting the Fermi level at the position of (c) W$_1$ ($E_{\rm F}$+17~meV), (d) W$_2$ ($E_{\rm F}$+65~meV), and (e) Rashba-split states ($E_{\rm F}$-25~meV). (f) Eliashberg spectral function $\alpha^2F(\omega)$ and EPC strength $\lambda(\omega)$. Energy distribution of the superconducting gap $\Delta_{\bf k}$ as a function of temperature calculated at (g) the Fermi level and by shifting the Fermi level at the position of (h) W$_1$ ($E_{\rm F}$+17~meV), (i) W$_2$ ($E_{\rm F}$+65~meV) and (j) Rashba-split states ($E_{\rm F}$-25~meV). All calculations were performed for the T$_d$ phase at the experimental lattice constants at ambient pressure. The  position, chirality and energy isosurface of the Weyl points were obtained with the WannierTools package~\cite{WANNTOOL}.}
\end{figure*}

Several studies have pointed out that the electronic structure of MoTe$_2$ is highly sensitive to the crystal structure parameters~\cite{RHODES, DENG, SAKANO, THIRUPATHAIAH, SI}, therefore it is important to examine the sensitivity of our results to small changes in the lattice constants. To this aim, we calculated the superconducting gap at the experimental unit cell parameters at ambient pressure while allowing the atomic positions to relax~\cite{SM}.  The resulting electronic structure and phonon dispersion are shown in Figs.~\ref{fig2}(h-n) and Figs.~\ref{fig3}(g-l). While the phonon frequencies are largely unaffected, we observe that the hole and electron bands shift down/up in energy relative to $E_{\rm F}$, respectively, by as much as 70~meV. These changes in the FS topology lead to a markedly narrower distribution of the superconducting gap (see Figs.~\ref{fig6}(c, f)). This time, although the gaps $\Delta_1$ and $\Delta_2$ associated with the electron and hole FS sheets overlap in energy, we can see two peaks at $\Delta_1=$~0.23~meV and $\Delta_2=$~0.20~meV in T$_d$, and $\Delta_1=$~0.21~meV and $\Delta_2=$~0.13~meV in 1T$^\prime$, respectively. These results are in line with magnetic penetration depth measurements of MoTe$_2$ at 0.45~GPa~\cite{GUGUCHIA} and specific heat measurements of S-doped MoTe$_{2-x}$S ($x \sim 0.2$)~\cite{LI}.

The numerical solutions to the ME equations give $T_{\rm c}$ values that are in an overall good agreement with the experimental measurements as shown in  Fig.~\ref{fig7}. While the critical temperature in the T$_d$ phase remains nearly constant under compression, in the 1T$^\prime$ phase it shows an almost four-fold increase, yielding a maximum value of 7.6~K at 8~GPa. This  increase in $T_{\rm c}$  provides evidence for a phonon-driven superconducting mechanism, and confirms that the 1T$^\prime$ phase is responsible for the sharp increase in the critical temperature observed experimentally at moderate pressures. By further analyzing the variation of the $T_{\rm c}$  and  EPC strength $\lambda$ with pressure (see Figs.~\ref{fig4}(c, d)), we find that  the dome-like shape superconducting phase diagram of MoTe$_2$ comes from the synergistic contribution of the density of state at the Fermi level $N_{\rm F}$ and the TA mode in the 1T$^\prime$ phase.

We further explore how to take advantage of the Weyl and Rashba-split states in the T$_d$ phase to achieve a superconducting system with nontrivial FS topology.  At the experimental lattice constants, the Weyl nodes are found at the intersection of electron and hole pockets 17~meV (W$_1$) and 65~meV (W$_2$) above the $E_{\rm F}$ (Fig.~\ref{fig8}(a,b)), in agreement with previous studies~\cite{SUN,SAKANO,DISSANAYAKE}. As shown in Fig.~\ref{fig8}(c-e), moving  $E_{\rm F}$ at the Weyl points or inside the Rashba split bands changes the Fermi surface, with the most substantial effect at the W$_2$ point where the inner hole and electron pocket FS2 and FS4 disappear and appears, respectively. As a result, for the W$_2$ point, there is an almost 10\% enhancement in the EPC strength due to the increased coupling of the high-energy Mo modes and a wider distribution of the superconducting gap which resides almost entirely on the electron pockets FS3 and FS4 as shown in Fig.~\ref{fig8}(f-j). Considering that the four W$_2$ points occur in pairs with opposite chirality, the electron-doped system provides the necessary conduction for realizing non-trivial sign-changing $s_{+-}$ pairing~\cite{HOSUR}.

\section{DISCUSSION}

We now comment on the difference between the calculated and measured critical temperature. Despite an overall good agreement between the theoretical and experimental results as shown in Fig.~\ref{fig7}, the $T_{\rm c}$, in particular at 0~GPa, is largely overestimated. In fact, an important question that remains open is why first-principles calculations systematically tend to overestimate the critical temperature and zero-temperature superconducting gap in transition metal dichalcogenides ~\cite{BEKAERT19,LEROUX,HEIL,ROSNER}.  In MoTe$_2$, we find that the critical temperature is reduced by about a factor of two if we consider the superconducting gap calculated at the experimental lattice parameters. In this case, we obtain $T_{\rm c}$ values of 1.7~K for T$_d$ and 1.6~K for 1T$^\prime$, which compare well with the 1.5~K and 1.33~K values at 0~GPa found in previous theoretical calculations~\cite{HEIKES}.  These results highlight the fact that small variations in the lattice parameters lead to appreciable changes in the electronic structure near the Fermi level which can then substantially affect the superconducting gap and critical temperature.  Additional calculations in Supplemental Fig.~S10~\cite{SM} also show how the variation in $N_{\rm F}$ can have a significant impact on the calculated $T_{\rm c}$ due to its direct effect on the EPC strength $\lambda$. However, all these differences alone cannot entirely account for the observed discrepancy with experiment.

Strong electron-electron interactions and spin fluctuations have been shown to considerably reduce the $T_{\rm c}$ when included alongside the electron-phonon interactions. For instance, for doped MoS$_2$ thin flake and bulk TiSe$_2$, the $T_{\rm c}$ was reduced by about a factor of two~\cite{DAS} and for FeB$_4$ by a factor of 25~\cite{BEKAERT18} to reach the experimental values. To investigate the effect of an enhanced Coulomb repulsion between the Cooper-pair electrons, we calculated the dependence of $T_{\rm c}$ on the Coulomb pseudopotential  $\mu^*$ as shown in Supplemental Fig.~S11. This analysis shows that to match the experimental $T_{\rm c}$ very different values of $\mu^*$ are required in the low, central, and high pressure regions. For instance,  a value  $\geq$ 0.3 is estimated at 0~GPa and the experimental lattice parameters at ambient pressure, well beyond the common range of 0.1-0.2. These results highlight the limitation of using a single constant $\mu^*$ parameter which could be overcome by computing the state-dependent electron-electron repulsion, as done for example in Refs.~[\onlinecite{HEIL,MARGINE16,ERREA,SANNA18}].

Finally, the compilation of the experimental $T_{\rm c}$ data in Fig.~\ref{fig7} illustrates that: (i) the values reported in various studies for the low pressure region can differ by as much as a factor of 3, and (ii) there are substantial differences observed between the onset and zero resistance $T_{\rm c}$ (see for instance the data from Ref.~[\onlinecite{QI}] shown as orange circles). Nonhydrostatic pressure conditions in the experimental setup, the coexistence of mixed phases, and the presence of phase inhomogeneities can in principle change the $T_{\rm c}$ considerably. This can explain, as for other superconductors~\cite{MARINI,LEROUX,WIENDLOCHA},  part of the discrepancy between different experiments as well as between experiments and theory. Therefore, further theoretical analysis combined with experimental measurements of superconducting properties are needed to better understand the dome-shaped superconducting diagram over the full pressure range.

\section{CONCLUSIONS}

We employed the {\it ab initio} anisotropic Migdal-Eliashberg theory to elucidate the nature of the superconducting paring mechanism in the T$_d$ and 1T$^\prime$ phases of MoTe$_2$. We show that the origin of the superconducting dome lies in the synergistic contribution of the density of states at the Fermi level and the TA mode frequency with pressure in the 1T$^\prime$ phase. Our calculations provide evidence for $s$-wave superconductivity in both phases, and the energy distribution of the order parameter is reminiscent of a two-gap scenario, although clearly disconnected superconducting gaps could not be identified. Based on our findings, the contribution of non-trivial states at the Weyl points can enhance the critical temperature, and possibly induce nonconventional pairing, in the non-centrosymmetric T$_d$ phase. This could be achieved by slightly moving up Fermi level via electron doping~\cite{LI,LI2,HOSUR}.

\section*{ACKNOWLEDGMENTS}
	
H.P. and E.R.M. acknowledge support from the National Science Foundation (Award No. OAC--1740263) and Binghamton University High Performance (SPIEDIE) Computing. This work used the Extreme Science and Engineering Discovery Environment (XSEDE)~\cite{XSEDE} which is supported by National Science Foundation grant number ACI-1548562. Specifically, this work used Comet at the San Diego Supercomputer Center through allocations TG--DMR180047 and TG--DMR180071. 
S.P. acknowledges support from the Leverhulme Trust (Grant No. RL-2012-001) and from the European Unions Horizon 2020
Research and Innovation Programme, under the Marie Sk\l{}odowska-Curie Grant Agreement SELPH2D No.~839217. 
Work by F.G. was supported by the U.S. Department of Energy (DOE), Office of
Science, Basic Energy Sciences (BES) under Award DE-SC0020129.


\begin{thebibliography}{10}

\bibitem{ALI}{M. N. Ali, J. Xiong, S. Flynn, J. Tao, Q. D. Gibson, L. M. Schoop, T. Liang, N. Haldolaarachchige, M. Hirschberger, N. P. Ong, and R. J. Cava, ``Large, non-saturating magnetoresistance in WTe$_2$", Nature \textbf{514}, 205 (2014).}

\bibitem{PAN}{X.C. Pan, X. Chen, H. Liu, Y. Feng, Z. Wei, Y. Zhou, Z. Chi, L. Pi, F. Yen, F. Song, and X. Wan, ``Pressure-driven dome-shaped superconductivity and electronic structural evolution in tungsten ditelluride", Nat. Commun. \textbf{6}, 7805 (2015)}

\bibitem{KANG}{D. Kang, Y. Zhou, W. Yi, C. Yang, J. Guo, Y. Shi, S. Zhang, Z. Wang, C. Zhang, S. Jiang, A. Li, K. Yang, Q. Wu, G. Zhang, L. Sun, and Z. Zhao, ``Superconductivity emerging from a suppressed large magnetoresistant state in tungsten ditelluride", Nat. Commun. \textbf{6}, 7804 (2015).}

\bibitem{KEUM}{D. H. Keum, S. Cho, J. H. Kim, D.-H. Choe, H.-J. Sung, M. Kan, H. Kang, J.-Y. Hwang, S. W. Kim, H. Yang, K. J. Chang, and Y. H. Lee, ``Bandgap opening in few-layered monoclinic MoTe$_2$", Nat. Phys. \textbf{11}, 482 (2015). }

\bibitem{LEE}{S. Lee, J. Jang, S.-I. Kim, S.-G. Jung, J. Kim, S. Cho, S. W. Kim, J. Y. Rhee, K.-S. Park, and T. Park, ``Origin of extremely large magnetoresistance in the candidate type-II Weyl semimetal MoTe$_{2-x}$", Sci. Rep. \textbf{8}, 13937 (2018).}

\bibitem{THIRUPATHAIAH}{S. Thirupathaiah, R. Jha, B. Pal, J. S. Matias, P. K. Das, P. K. Sivakumar, I. Vobornik, N. C. Plumb, M. Shi, R. A. Ribeiro, and D. D. Sarma, ``MoTe$_2$: An uncompensated semimetal with extremely large magnetoresistance", Phys. Rev. B \textbf{95}, 241105(R) (2017).}

\bibitem{QI}{Y. Qi, P. G. Naumov, M. N. Ali, C. R. Rajamathi, O. Barkalov, Y. Sun, C. Shekhar, S.-C. Wu, V. S\"{u}\ss, M. Schmidt, E. Pippel, P. Werner, R. Hillebrand, T. F\"{o}rster, E. Kampertt, W. Schnelle, S. Parkin, R. J. Cava, C. Felser, B. Yan, and S. A. Medvedev, ``Superconductivity in Weyl semimetal candidate MoTe$_2$", Nat. Commun. \textbf{7}, 11038 (2016).}

\bibitem{TAKAHASHI}{H. Takahashi, T. Akiba, K. Imura, T. Shiino, K. Deguchi, N. K. Sato, H. Sakai, M. S. Bahramy, and S. Ishiwata, ``Anticorrelation between polar lattice instability and superconductivity in the Weyl semimetal candidate MoTe$_2$", Phys. Rev. B \textbf{95}, 1005001(R) (2017).}

\bibitem{HEIKES}{C. Heikes, I. L. Liu, T. Metz, C. Eckberg, P. Neves, Y. Wu, L. Hung, P. Piccoli, H. Cao, J. Leao, J. Paglione, T. Yildirim, N. P. Butch, and W. Ratcliff, ``Mechanical control of crystal symmetry and superconductivity in Weyl semimetal MoTe$_2$", Phys. Rev. Mater. \textbf{2}, 074202  (2018).}

\bibitem{GUGUCHIA}{Z. Guguchia, F. von Rohr, Z. Shermadini, A. T. Lee, S. Banerjee, A. R. Wieteska, C. A. Marianetti, H. Luetkens, Z. Gong, B. A. Frandsen, S. C. Cheung, C. Baines, A. Shengelaya, A. N. Pasupathy, E. Morenzoni, S. J. L. Billinge, A. Amato, R. J. Cava, R. Khasanov, and Y. J. Uemura, ``Signatures of the topological s$^{+-}$ superconducting order parameter in the type-II Weyl semimetal T$_d$-MoTe$_2$", Nat. Commun. \textbf{8}, 1082 (2017).}

\bibitem{DISSANAYAKE}{S. Dissanayake, C. Duan, J. Yang,  J. Liu, M. Matsuda, C. Yue, J. A. Schneeloch, J. C. Y. Teo, and D. Louca, ``Electronic band tuning under pressure in MoTe$_2$ topological semimetal", NPJ Quantum Mater. \textbf{4}, 45 (2019).}

\bibitem{SUN}{Y. Sun, S. C. Wu, M. N. Ali, C. Felser, and B. Yan, ``Prediction of Weyl semimetal in orthorhombic MoTe$_2$", Phys. Rev. B. \textbf{92}, 161107(R) (2015).}

\bibitem{SOLUYANOV}{A. A. Soluyanov, D. Gresch, Z. Wang, Q. Wu, M. Troyer, X. Dai, and B. A. Bernevig, ``Type-II Weyl semimetals", Nature \textbf{527}, 495–498 (2015).}

\bibitem{WANG}{Z. Wang, D. Gresch, A. A. Soluyanov, W. Xie, S. Kushwaha, X. Dai, M. Troyer, R. J. Cava, and B. A. Bernevig, ``MoTe$_2$: A type-II Weyl topological metal", Phys. Rev. Lett. \textbf{117}, 056805 (2016).}

\bibitem{TAMAI}{A. Tamai, Q. S. Wu, I. Cucchi, F. Y. Bruno, S. Ricc\`{o}, T. K. Kim, M. Hoesch, C. Barreteau, E. Giannini, C. Besnard, A. A. Soluyanov, and F. Baumberger, ``Fermi Arcs and Their Topological Character in the Candidate Type-II Weyl Semimetal MoTe$_2$", Phys. Rev. X \textbf{6}, 031021 (2016).}

\bibitem{DENG}{K. Deng, G. Wan, P. Deng, K. Zhang, S. Ding, E. Wang, M. Yan, H. Huang, H. Zhang, Z. Xu, J. Denlinger, A. Fedorov, H. Yang, W. Duan, H. Yao, Y. Wu, S. Fan, H. Zhang, X. Chen, and S. Zhou, ``Experimental observation of topological Fermi arcs in type-II Weyl semimetal MoTe$_2$", Nat. Phys. \textbf{12}, 1105 (2016).}

\bibitem{LI}{Y. Li, Q. Gu, C. Chen, J. Zhang, Q. Liu, X. Hu, J. Liu, Y. Liu, L. Ling, M. Tian, Y. Wang, N. Samarth, S. Li, T. Zhang, J. Feng, and J. Wang, ``Nontrivial superconductivity in topological MoTe$_{2-x}$S$_x$ crystals", PNAS \textbf{115}, 9503 (2018).}

\bibitem{LUO}{X. Luo, F. C. Chen, J. L. Zhang, Q. L. Pei, G. T. Lin, W. J. Lu, Y. Y. Han, C. Y. Xi,
W. H. Song, and Y. P. Sun, ``T$_d$-MoTe$_2$: A possible topological superconductor", Appl. Phys. Lett. \textbf{109}, 102601 (2016).}

\bibitem{NAIDYUK}{Y. Naidyuk, O. Kvitnitskaya, D. Bashlakov, S. Aswartham, I. Morozov, I. Chernyavskii, G. Fuchs, S.-L. Drechsler, R. H\"{u}hne, K. Nielsch, B. B\"{u}chner, and D. Efremov, ``Surface superconductivity in the Weyl semimetal MoTe$_2$ detected by point contact spectroscopy", 2D Mater. \textbf{5},  045014 (2018).} 

\bibitem{HOSUR}{P. Hosur, X. Dai, Z. Fang, and X.-L. Qi, ``Time-reversal-invariant topological superconductivity in doped Weyl semimetals", Phys. Rev. B \textbf{90}, 045130 (2014).}

\bibitem{CHEN}{F. C. Chen, X. Luo, R. C. Xiao, W. J. Lu, B. Zhang, H. X. Yang, J. Q. Li, Q. L. Pei, D. F. Shao, R. R. Zhang, L. S. Ling, C. Y. Xi, W. H. Song, and Y. P. Sun, ``Superconductivity enhancement in the S-doped Weyl semimetal candidate MoTe$_2$", Appl. Phys. Lett. \textbf{108}, 162601 (2016).}

\bibitem{MANDAL}{M. Mandal, S. Marik, K. P. Sajilesh, Arushi, D. Singh, J. Chakraborty, N. Ganguli, and R. P. Singh, ``Enhancement of the superconducting transition temperature by Re doping in Weyl semimetal MoTe$_2$", Phys. Rev. Materials \textbf{2}, 094201 (2018).}

\bibitem{CHO}{S. Cho, S.H. Kang, H.S. Yu, H.W. Kim, W. Ko, S.W. Hwang, W.H. Han, D.H. Choe, Y.H. Jung, K.J. Chang, and Y.H. Lee, ``Te vacancy-driven superconductivity in orthorhombic molybdenum ditelluride", 2D Mater. \textbf{4}, 021030 (2017).}

\bibitem{LI2}{P. Li, J. Cui, J. Zhou,  D. Guo, J. Yi, Z. Zhao, J. Fan, Z. Ji, X. Jing, F. Qu, C. Yang, L. Lu, J. Lin,  Z. Liu, and G. Liu, ``Phase transition and superconductivity enhancement in Se-substituted MoTe$_2$ thin films", Adv. Mater. \textbf{31}, 1904641 (2019). }

\bibitem{ALLEN1}{P. B. Allen and B. Mitrovi\'{c}, ``Theory of superconducting T$_c$", Solid State Phys. \textbf{37}, 1 (1982).}

\bibitem{MARGINE1}{E. R. Margine and F. Giustino, ``Anisotropic Migdal-Eliashberg theory using Wannier functions", Phys. Rev. B \textbf{87}, 024505 (2013).}


\bibitem{QE}{P. Giannozzi, O. Andreussi, T. Brumme, O. Bunau, M. B. Nardelli, M. Calandra, R. Car, C. Cavazzoni, D. Ceresoli, M. Cococcioni  {\it et al.}, ``Advanced capabilities for materials modelling with Quantum ESPRESSO", J. Phys.: Condens. Matter \textbf{29}, 465901 (2017).}

\bibitem{NC-PP}{E. Kucukbenli, M. Monni, B. I. Adetunji, X. Ge, G. A. Adebayo, N. Marzari, S. De Gironcoli, and A. Dal Corso. ``Projector augmented-wave and all-electron calculations across the periodic table: a comparison of structural and energetic properties." arXiv preprint \textbf{arXiv:1404.3015} (2014).}

\bibitem{PBE}{J. P. Perdew, K. Burke, and M. Ernzerhof, ``Generalized gradient approximation made simple", Phys. Rev. Lett. \textbf{77}, 3865 (1996).}

\bibitem{optB86b}{J. Klime\v{s}, D. R. Bowler, and A. Michaelides, ``Van der Waals density functionals applied to solids", Phys. Rev. B \textbf{83}, 195131 (2011).}

\bibitem{vdW}{J. Klime\v{s}, D. R. Bowler, and A. Michaelides, ``Chemical accuracy for the van der Waals density functional", J. Phys.: Condens. Matter \textbf{22}, 022201 (2010).}

\bibitem{k-mesh}{H. J. Monkhorst and J. D. Pack, ``Special points for Brillouin-zone integrations", Phys. Rev. B \textbf{13}, 5188 (1976).}

\bibitem{smearing}{M. Methfessel and A. T. Paxton, ``High-precision sampling for Brillouin-zone integration in metals", Phys. Rev. B \textbf{40}, 3616 (1989).}

\bibitem{DFPT}{S. Baroni, S. de Gironcoli, A. Dal Corso, and P. Giannozzi, ``Phonons and related properties of extended systems from density-functional perturbation theory", Rev. Mod. Phys. \textbf{73}, 515 (2001).}

\bibitem{Giustino2007}{F. Giustino, M. L. Cohen, and S. G. Louie, ``Electron-phonon interaction using Wannier functions", Phys. Rev. B \textbf{76}, 165108 (2007).}

\bibitem{EPW}{S. Ponc\'{e}, E. R. Margine, C. Verdi, and F. Giustino, ``EPW: Electron-phonon coupling, transport and superconducting properties using maximally localized Wannier functions", Comput. Phys. Commun. \textbf{209}, 116 (2016).}

\bibitem{WANN1}{N. Marzari, A. A. Mostofi, J. R. Yates, I. Souza, and D. Vanderbilt, ``Maximally localized Wannier functions: Theory and applications", Rev. Mod. Phys. \textbf{84}, 1419 (2012).}


\bibitem{WANN2}{G. Pizzi, V. Vitale, R. Arita, S. Bluegel, F. Freimuth, G. G\'eranton, M. Gibertini, D. Gresch, C. Johnson, T. Koretsune {\it et al.}, ``Wannier90 as a community code: new features and applications", J. Phys.: Condens. Matter. \textbf{32(16)}, 165902 (2019).}

\bibitem{BROWN}{B. E. Brown, ``The Crystal Structures of WTe$_2$ and High-Temperature MoTe$_2$", Acta Cryst. \textbf{20}, 268 (1966).}

\bibitem{SM}{See Supplemental Material at [url] for further details on the first-principles calculations, Figs. S1-S11 and Table S1.}

\bibitem{ZHANG}{K. Zhang, C. Bao, Q. Gu, X. Ren, H. Zhang, K. Deng, Y. Wu, Y. Li, J. Feng, and S. Zhou, ``Raman signatures of inversion symmetry breaking and structural phase transition in type-II Weyl semimetal MoTe$_2$", Nat. Comm. \textbf{7}, 13552 (2016).}

\bibitem{SONG}{Q. Song, H. Wang, X. Pan, X. Xu, Y. Wang, Y. Li, F. Song, X. Wan, Y. Ye, and L. Dai, ``Anomalous in-plane anisotropic Raman response of monoclinic semimetal 1T$^\prime$-MoTe$_2$", Sci. Rep. \textbf{7}, 1758 (2017).}

\bibitem{OLIVER}{S. M. Oliver, R. Beams, S. Krylyuk, I. Kalish, A. K. Singh , A. Bruma, F. Tavazza, J. Joshi, I. R. Stone, S. J. Stranick, A. V Davydov, and P. M. Vora, ``The structural phases and vibrational properties of Mo$_{1-x}$W$_x$Te$_2$ alloys", 2D Mater. \textbf{4}, 045008 (2017).}

\bibitem{CHEN2}{S-Y. Chen, T. Goldstein, D. Venkataraman, A. Ramasubramaniam, and J. Yan, ``Activation of new Raman modes by inversion symmetry breaking in type II Weyl semimetal candidate T$^\prime$-MoTe$_2$", Nano Lett. \textbf{16},  5852 (2016).}

\bibitem{SI}{J. G. Si, W. J. Lu, H. Y. Lv, B. C. Zhao, and Y. P. Sun, ``Pressure controllable phase transition in MoTe$_2$ by the interlayer band occupancy", Phys. Lett. A \textbf{383}, 126016 (2019).}

\bibitem{LU}{P. Lu, J.-S. Kim, J. Yang, H. Gao, J. Wu, D. Shao, B. Li, D. Zhou, J. Sun, D. Akinwande, D. Xing, and J.-F. Lin, ``Origin of superconductivity in the Weyl semimetal WTe$_2$ under pressure", Phys. Rev. B \textbf{94}, 224512 (2016).}

\bibitem{VESTA}{K. Momma and F. Izumi, ``VESTA 3 for three-dimensional visualization of crystal, volumetric and morphology data," J. Appl. Crystallogr. \textbf{44}, 1272-1276 (2011).}

\bibitem{BCS}{J. Bardeen, L. N. Cooper, and J. R. Schrieffer, ``Theory of Superconductivity", Phys. Rev. \textbf{108}, 1175 (1957).}

\bibitem{RHODES}{D. Rhodes, R. Sch\"{o}nemann, N. Aryal, Q. Zhou, Q. R. Zhang, E. Kampert, Y.-C. Chiu, Y. Lai, Y. Shimura, G. T. McCandless, J. Y. Chan, D. W. Paley, J. Lee, A. D. Finke, J. P. C. Ruff, S. Das, E. Manousakis, and L. Balicas, ``Bulk Fermi surface of the Weyl type-II semimetallic candidate $\gamma$-MoTe$_2$", Phys. Rev. B \textbf{96}, 165134 (2017).}

\bibitem{SAKANO}{M. Sakano, M. S. Bahramy, H. Tsuji, I. Araya, K. Ikeura, H. Sakai, S. Ishiwata, K. Yaji, K. Kuroda, A. Harasawa, S. Shin, and K. Ishizaka, ``Observation of spin-polarized bands and domain-dependent Fermi arcs	in polar Weyl semimetal MoTe$_2$", Phys. Rev. B \textbf{95}, 121101(R) (2017).}

\bibitem{WANNTOOL}{Q. Wu, S. Zhang, H-F. Song, M. Troyer and A. A. Soluyanov, ``WannierTools: an open-source software package for novel topological materials", Comput. Phys. Commun. \textbf{224}, 405 (2018).}

\bibitem{BEKAERT19}{J. Bekaert, E. Khestanova, D.G. Hopkinson, J. Birkbeck, N. Clark, M. Zhu, D.A. Bandurin, R. Gorbachev, S. Fairclough, Y. Zou, M. Hamer, D. J. Terry, J. J. P. Peters, A. M. Sanchez, B. Partoens, S. J. Haigh, M. V. Milo\v{s}evi\'{c}, and I. V. Grigorieva ``Enhanced superconductivity in few-layer TaS$_2$ due to healing by oxygenation", Nano. Lett. \textbf{20(5)}, 3808 (2020).}

\bibitem{LEROUX}{M. Leroux, I. Errea, M. L. Tacon, S-M. Souliou, G. Garbarino, L. Cario, A. Bosak, F. Mauri, M. Calandra, and P. Rodi\'{e}re, ``Strong anharmonicity induces quantum melting of charge density wave in 2\textit{H}-NbSe$_2$ under pressure", Phys. Rev. B \textbf{92}, 140303(R) (2015).}

\bibitem{HEIL}{C. Heil, S. Ponc\'{e}, H. Lambert, M. Schlipf, E. R. Margine, and F. Giustino, ``Origin of Superconductivity and Latent Charge Density Wave in NbS$_2$", Phys. Rev. Lett. \textbf{119}, 087003 (2017).}

\bibitem{ROSNER}{M. R\"{o}sner, S. Haas, and T. O. Wehling, ``Phase diagram of electron-doped dichalcogenides", Phys. Rev. B \textbf{90}, 245105 (2014).}

\bibitem{DAS}{T. Das and K. Dolui, ``Superconducting dome in MoS$_2$ and TiSe$_2$ generated by quasiparticle-phonon coupling", Phys. Rev. B \textbf{91}, 094510 (2015).}

\bibitem{BEKAERT18}{J. Bekaert, A. Aperis, B. Partoens, P. M. Oppeneer, and M. V. Milo\v{s}evi\'{c}, ``Advanced first-principles theory of superconductivity including both lattice vibrations and spin fluctuations: The case of FeB$_4$", Phys. Rev. B \textbf{97}, 014503 (2018).}

\bibitem{MARGINE16} {E. R. Margine, H. Lambert, and F. Giustino. ``Electron-phonon interaction and pairing mechanism in superconducting Ca-intercalated bilayer graphene", Sci. Rep. \textbf{6}, 21414 (2016).}

\bibitem{ERREA} {I. Errea, F. Belli, L. Monacelli, A. Sanna, T. Koretsune, T. Tadano, R. Bianco, M. Calandra, R. Arita, F. Mauri, and J. A. Flores-Livas, ``Quantum crystal structure in the 250-kelvin superconducting lanthanum hydride", Nature \textbf{578(7793)}, 66 (2020).}

\bibitem{SANNA18} {A. Sanna, J. A. Flores-Livas, A. Davydov, G. Profeta, K. Dewhurst, S. Sharma, and E. K. U. Gross, ``Ab initio Eliashberg Theory: Making Genuine Predictions of Superconducting Features", J. Phys. Soc. Jpn \textbf{87(4)}, 041012 (2018).}

\bibitem{MARINI}{G. Marini, P. Barone, A. Sanna, C. Tresca, L. Benfatto, and G. Profeta, ``Superconductivity in tin selenide under pressure", Phys. Rev. Mater. \textbf{3}, 114803 (2019).}

\bibitem{WIENDLOCHA}{B. Wiendlocha, R. Szcz\c{e}\'{s}niak, A. P. Durajski, and M. Muras, ``Pressure effects on the unconventional superconductivity of noncentrosymmetric LaNiC$_2$", Phys. Rev. B \textbf{94}, 134517 (2016).}

\bibitem{XSEDE}{J. Towns, T. Cockerill, M. Dahan, I. Foster, K. Gaither, A. Grimshaw, V. Hazlewood, S. Lathrop, D. Lifka, G. D. Peterson, R. Roskies, J. R. Scott, N. Wilkins-Diehr, ``XSEDE: Accelerating scientific discovery", Comput. Sci. {\&} Eng. \textbf{16}, 62-74 (2014).}

	
\end{thebibliography}
\end{document}


\title{Supplemental Material \\Superconducting properties of MoTe$_2$ from the \textit{ab initio} anisotropic Migdal-Eliashberg theory}
	
\author{Hari Paudyal}
\affiliation{Department of Physics, Applied Physics, and Astronomy, Binghamton University-SUNY, Binghamton, New York 13902, USA}
\author{Samuel Ponc\'{e}}
\affiliation{Department of Materials, University of Oxford, Parks Road, Oxford OX1 3PH, United Kingdom}
\affiliation{Theory and Simulation of Materials (THEOS), \'Ecole Polytechnique F\'ed\'erale de Lausanne, CH-1015 Lausanne, Switzerland}
\author{Feliciano Giustino}
\affiliation{Oden Institute for Computational Engineering and Sciences, The University of Texas at Austin, Austin, Texas 78712, USA}
\affiliation{Department of Physics, The University of Texas at Austin, Austin, Texas 78712, USA}
\author{Elena R. Margine}
\email{rmargine@binghamton.edu}
\affiliation{Department of Physics, Applied Physics, and Astronomy, Binghamton University-SUNY, Binghamton, New York 13902, USA}
	
\maketitle

\begin{figure}[h!]
\centering
\includegraphics[width=0.81\linewidth]{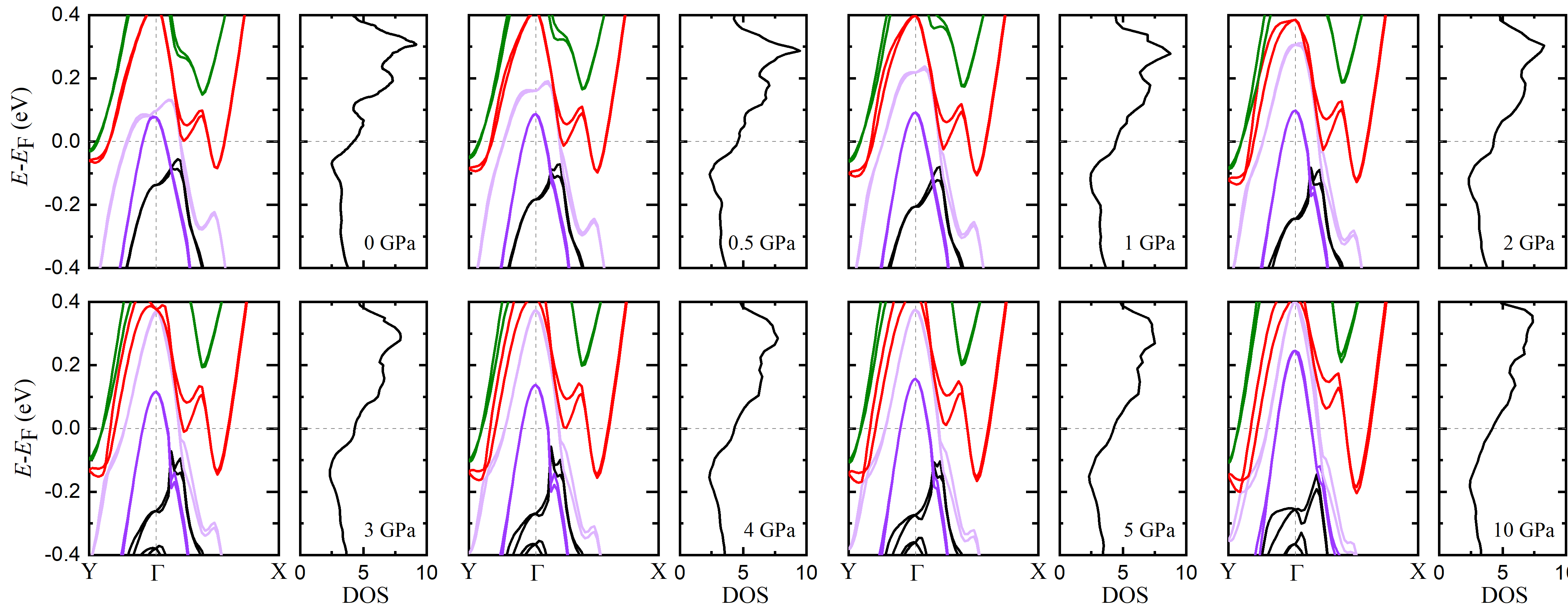} 
\caption{\label{fig-S1} Calculated electronic band structure and DOS (states/eV/u.c) for the T$_d$ phase at various pressures. A finite splitting of the bands can be observed in the T$_d$ phase due to the combined effect of SOC and lack of inversion symmetry (each pair of bands is shown with the same color).}
\end{figure}

\begin{figure}[h!]
\centering
\includegraphics[width=0.81\linewidth]{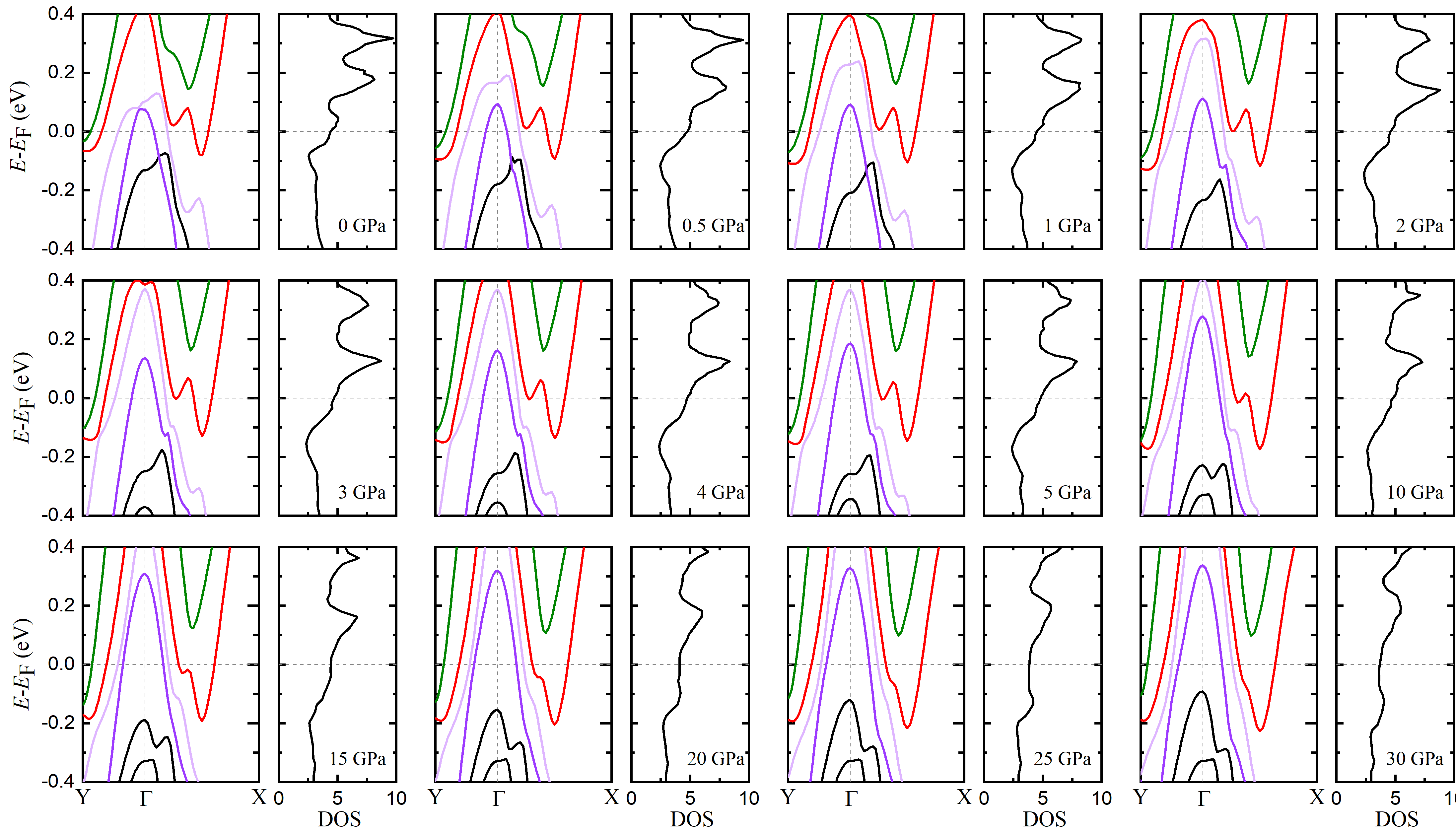}  
\caption{\label{fig-S2} Calculated electronic band structure and DOS (states/eV/u.c) for the 1T$^\prime$ phase at various pressures.}	
\end{figure}

\begin{figure}[h!]
	\centering
	\includegraphics[width=0.99\linewidth]{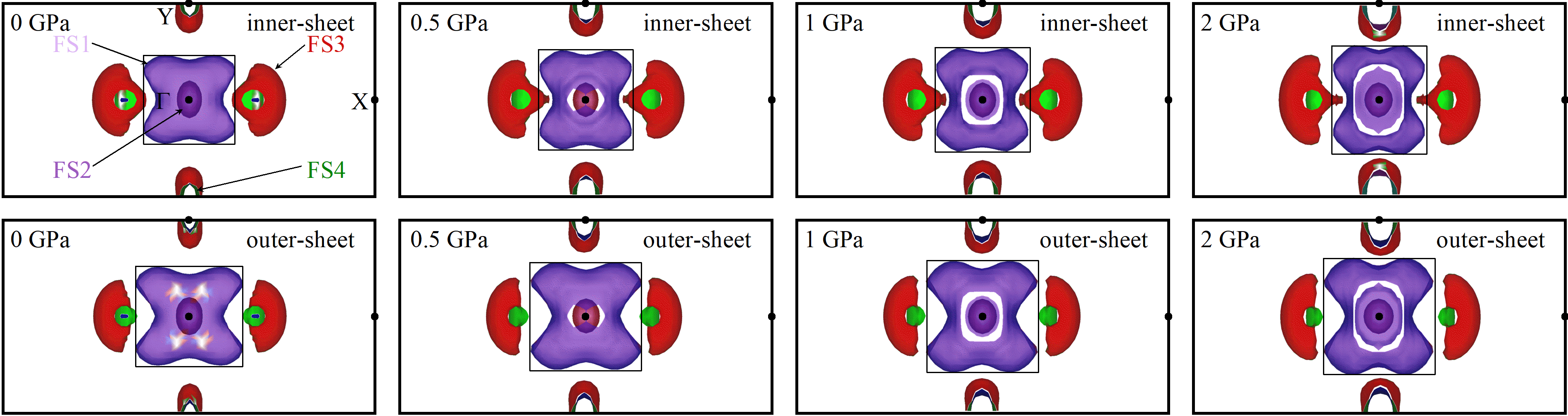}
    \vskip 0.2in
	\includegraphics[width=0.99\linewidth]{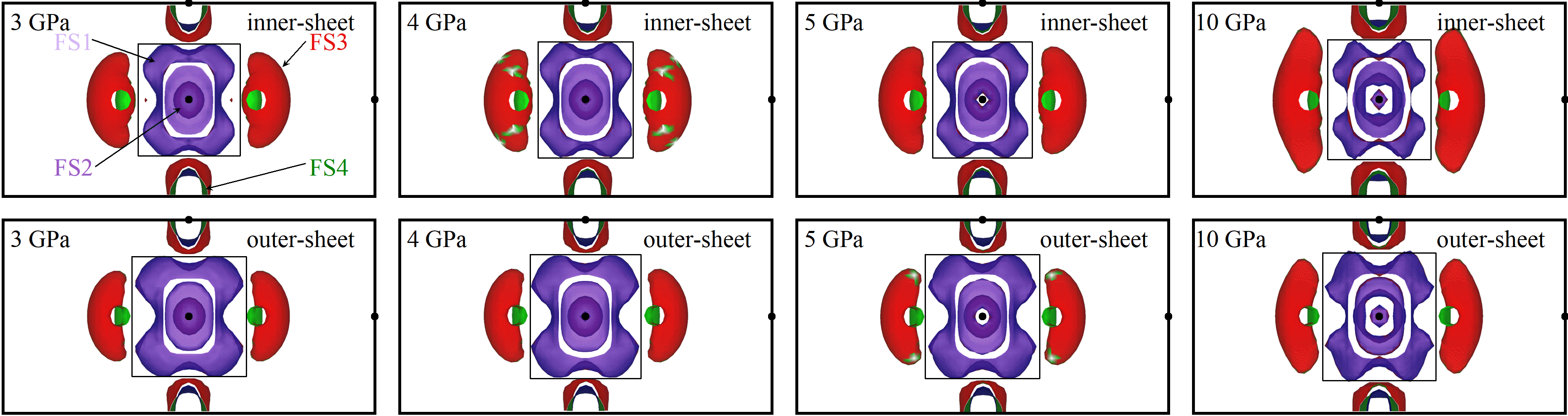} 
	\caption{\label{fig-S3} Top view of the FS for the T$_d$ phase at various pressures. Each pocket splits in a pair of inner and outer sheets  due to the combined effect of SOC and lack of inversion symmetry.  The inside and outside of FS3 are shown in light green and red colors, while the inside and outside of FS4 are shown in dark blue and dark green colors. The black rectangle marks the boundary between the regions corresponding to the electron and hole pockets.}
\end{figure}
\begin{figure}[h!]
	\centering
	\includegraphics[width=0.99\linewidth]{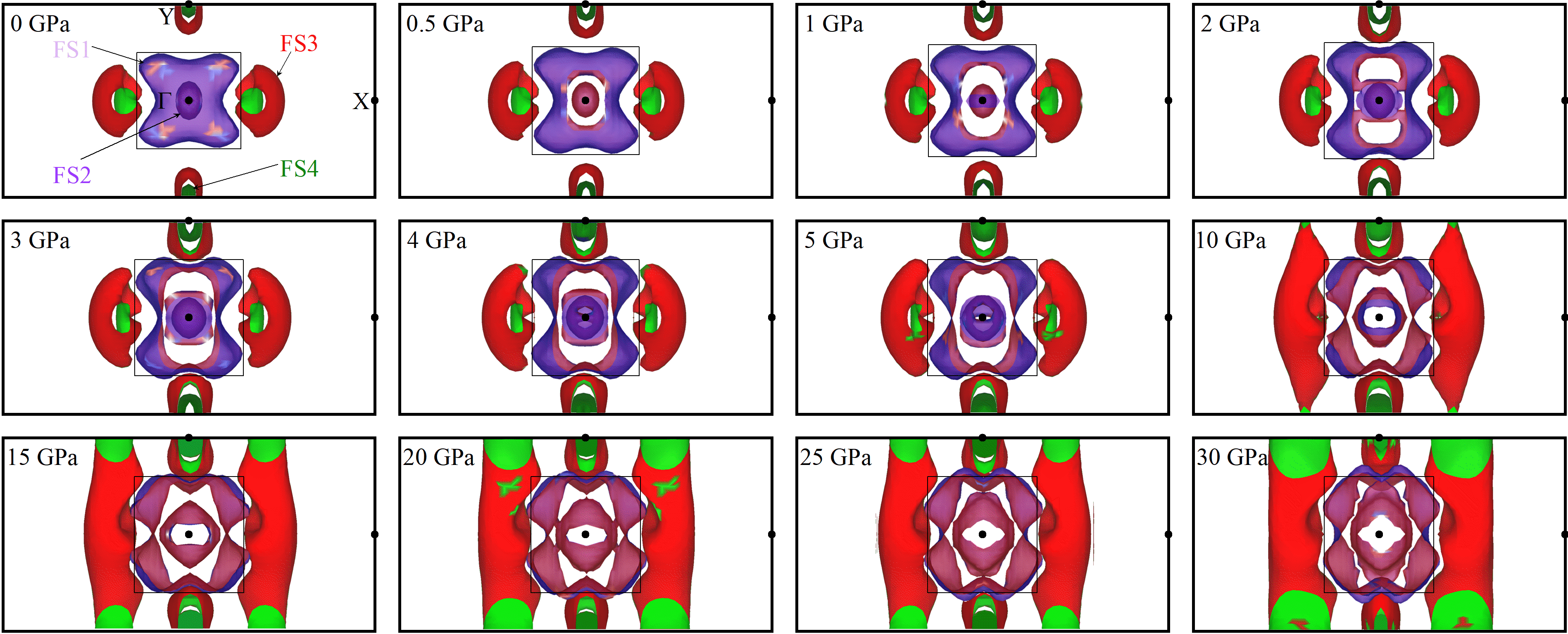}  
	\caption{\label{fig-S4} Top view of the FS for the 1T$^\prime$ phase at various pressures. The inside and outside of FS3 are shown in light green and red colors, while the inside and outside of FS4 are shown in dark blue and dark green colors. The black rectangle marks the boundary between the regions corresponding to the electron and hole pockets.}
\end{figure}

\begin{figure}[h!]
\centering
\includegraphics[width=0.495\linewidth]{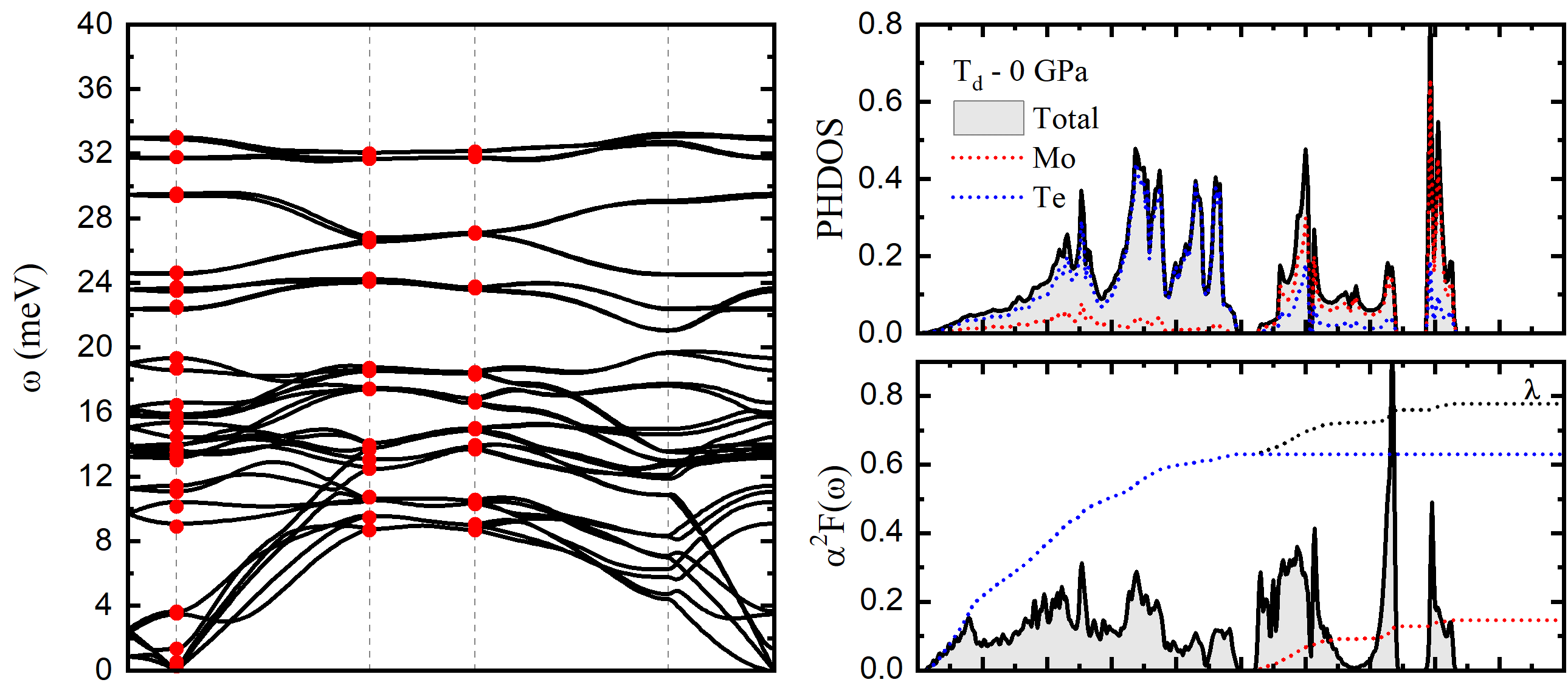}
\includegraphics[width=0.495\linewidth]{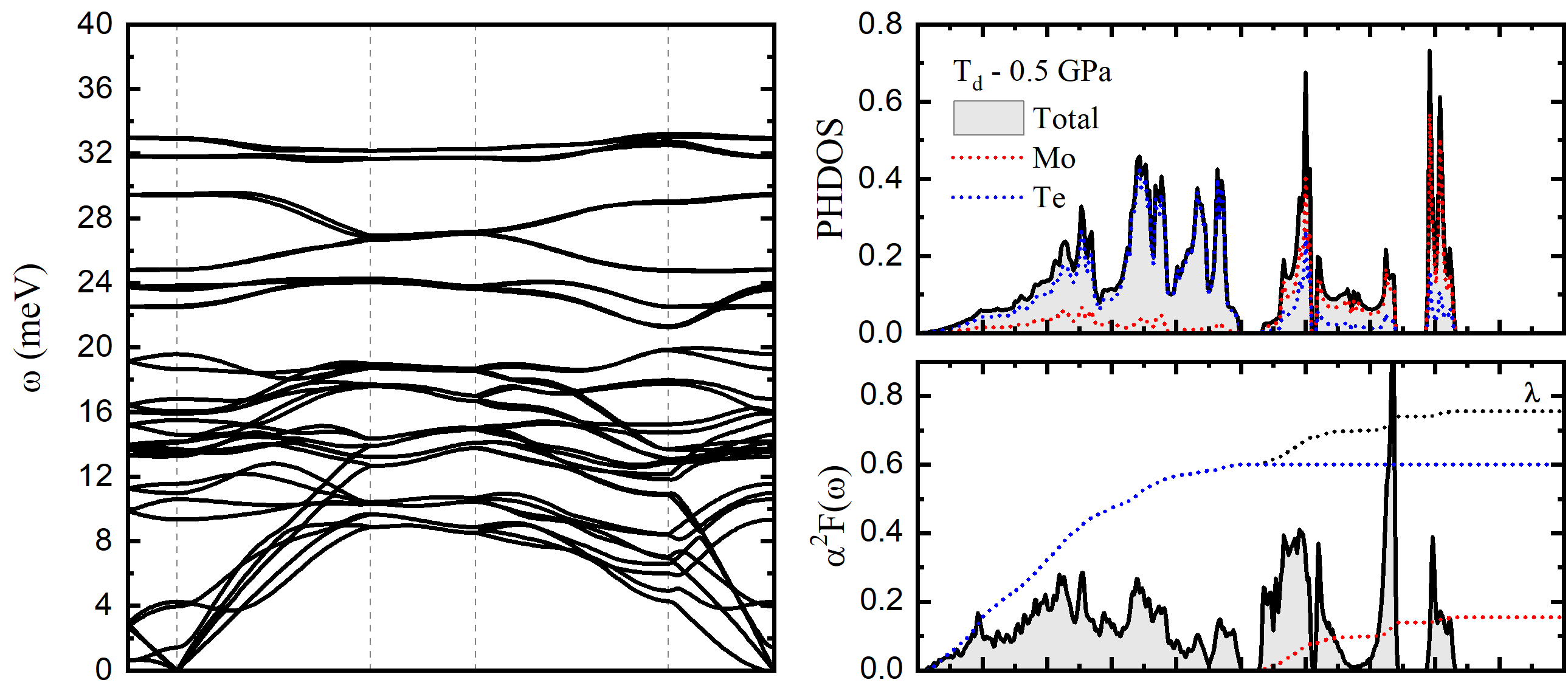}	
\includegraphics[width=0.495\linewidth]{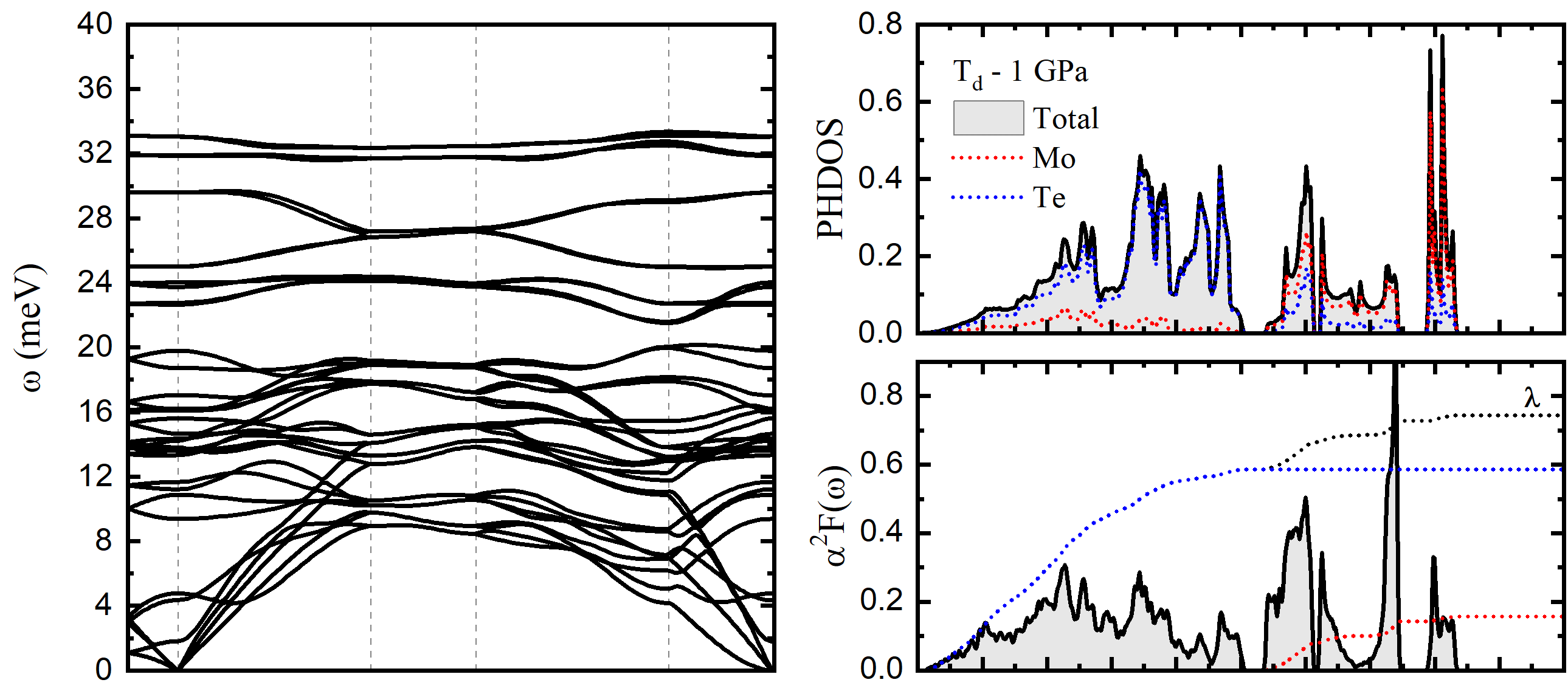}
\includegraphics[width=0.495\linewidth]{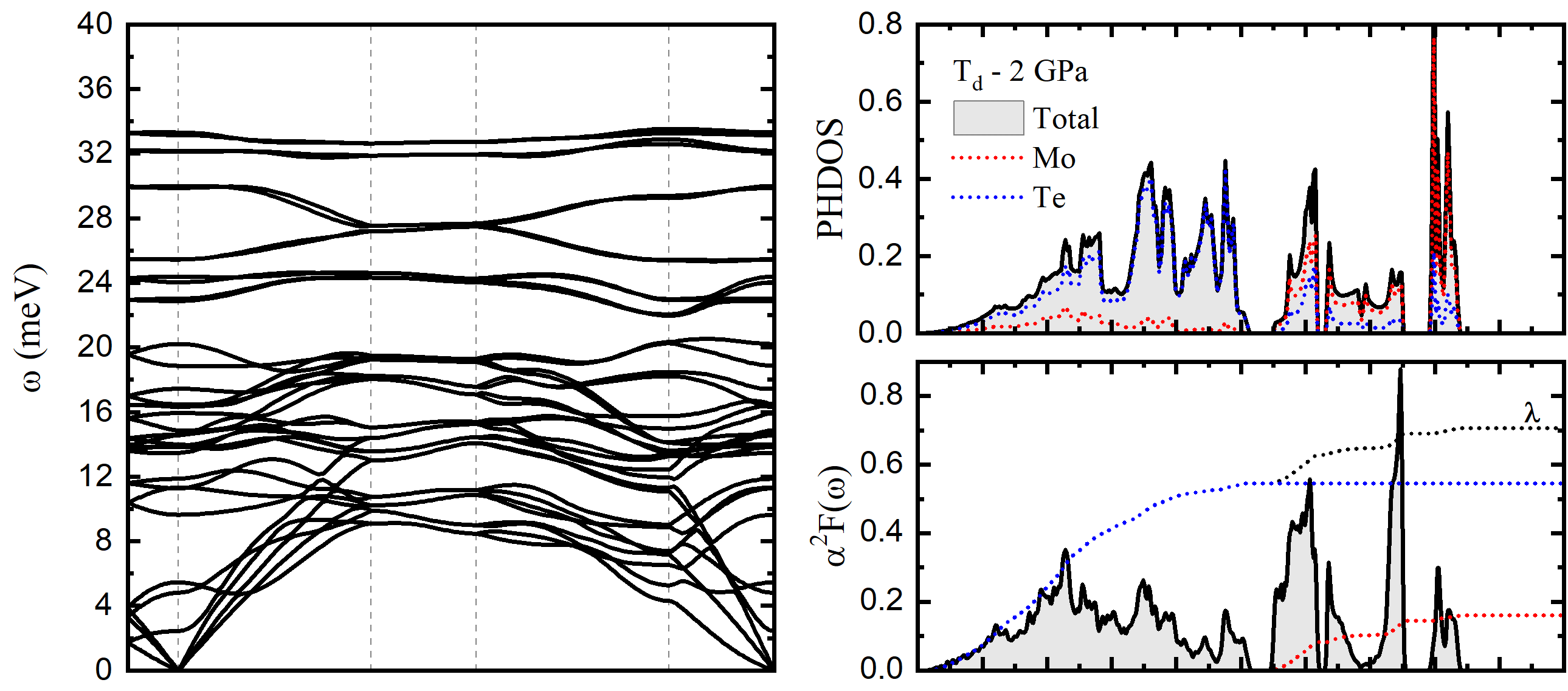}
\includegraphics[width=0.495\linewidth]{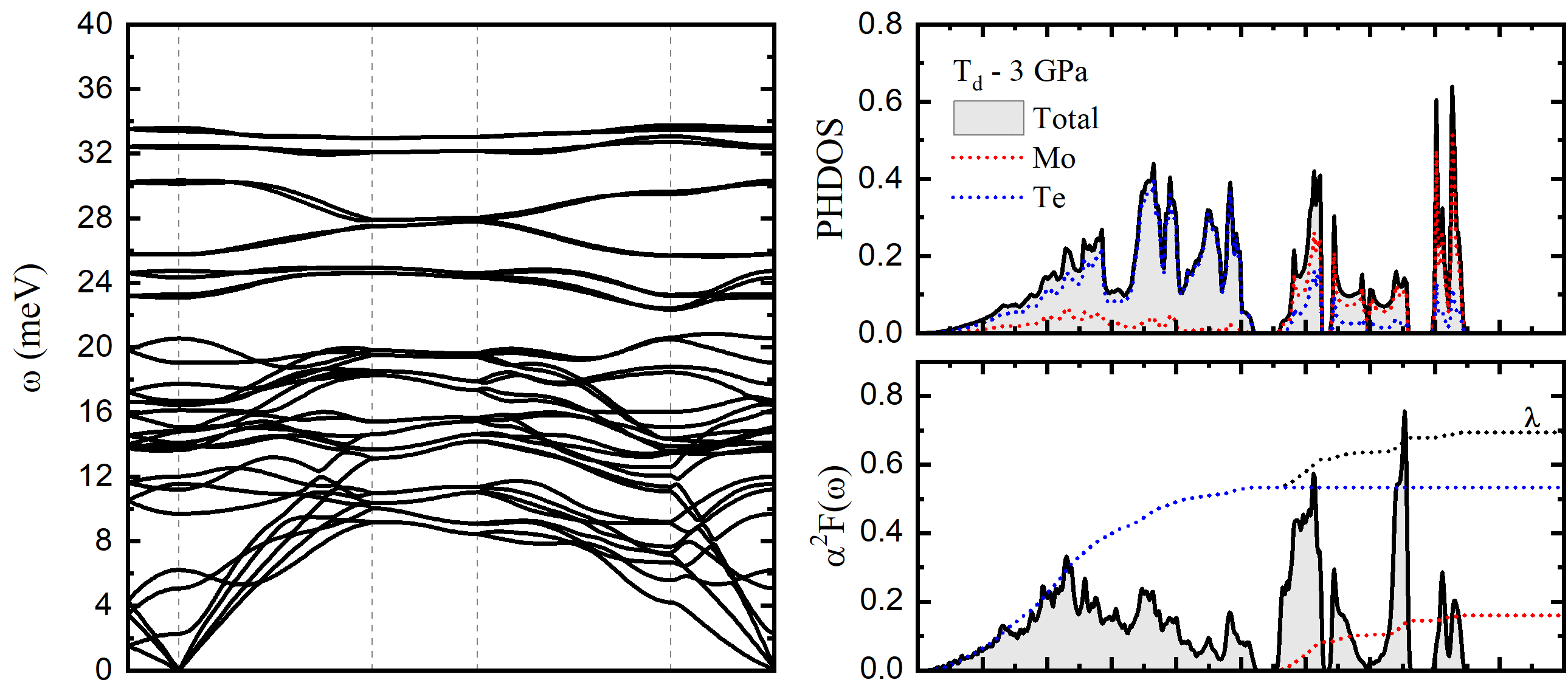}
\includegraphics[width=0.495\linewidth]{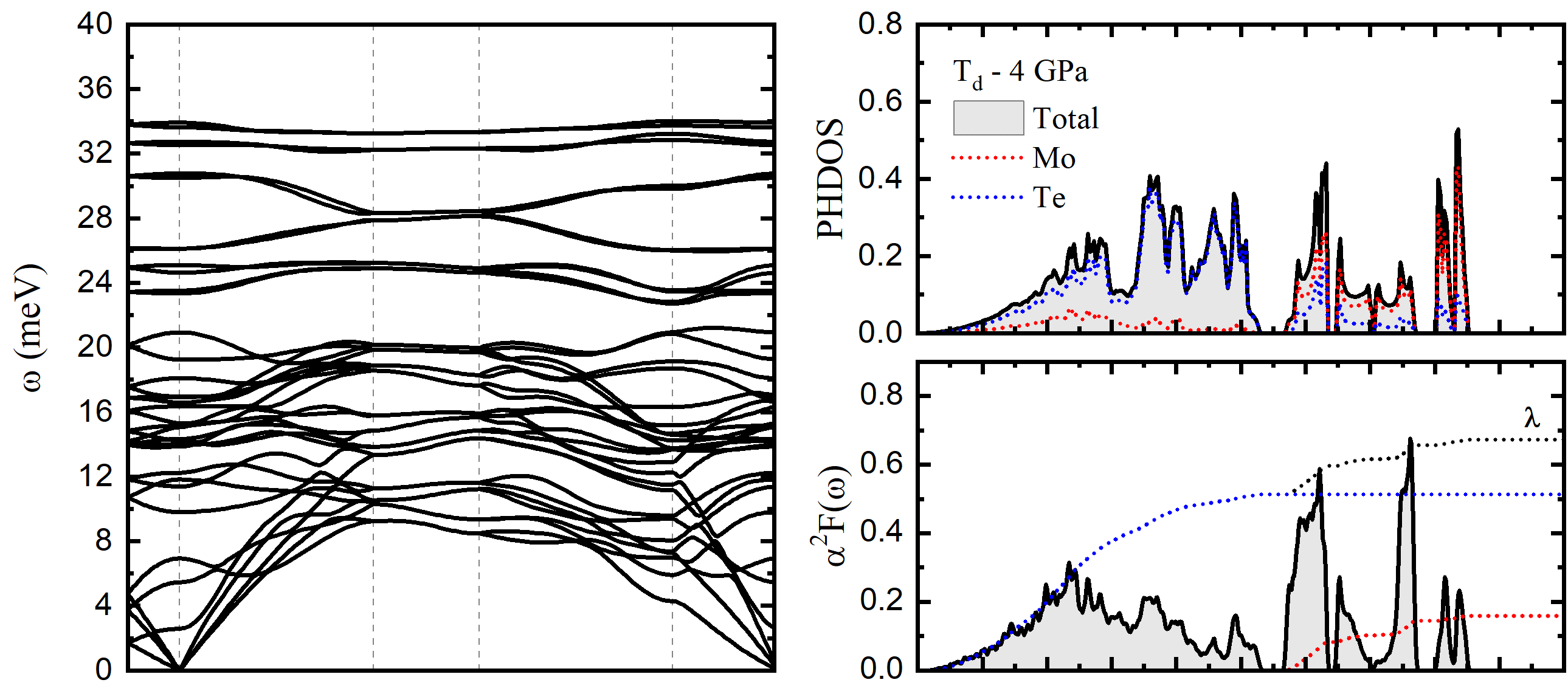}
\includegraphics[width=0.495\linewidth]{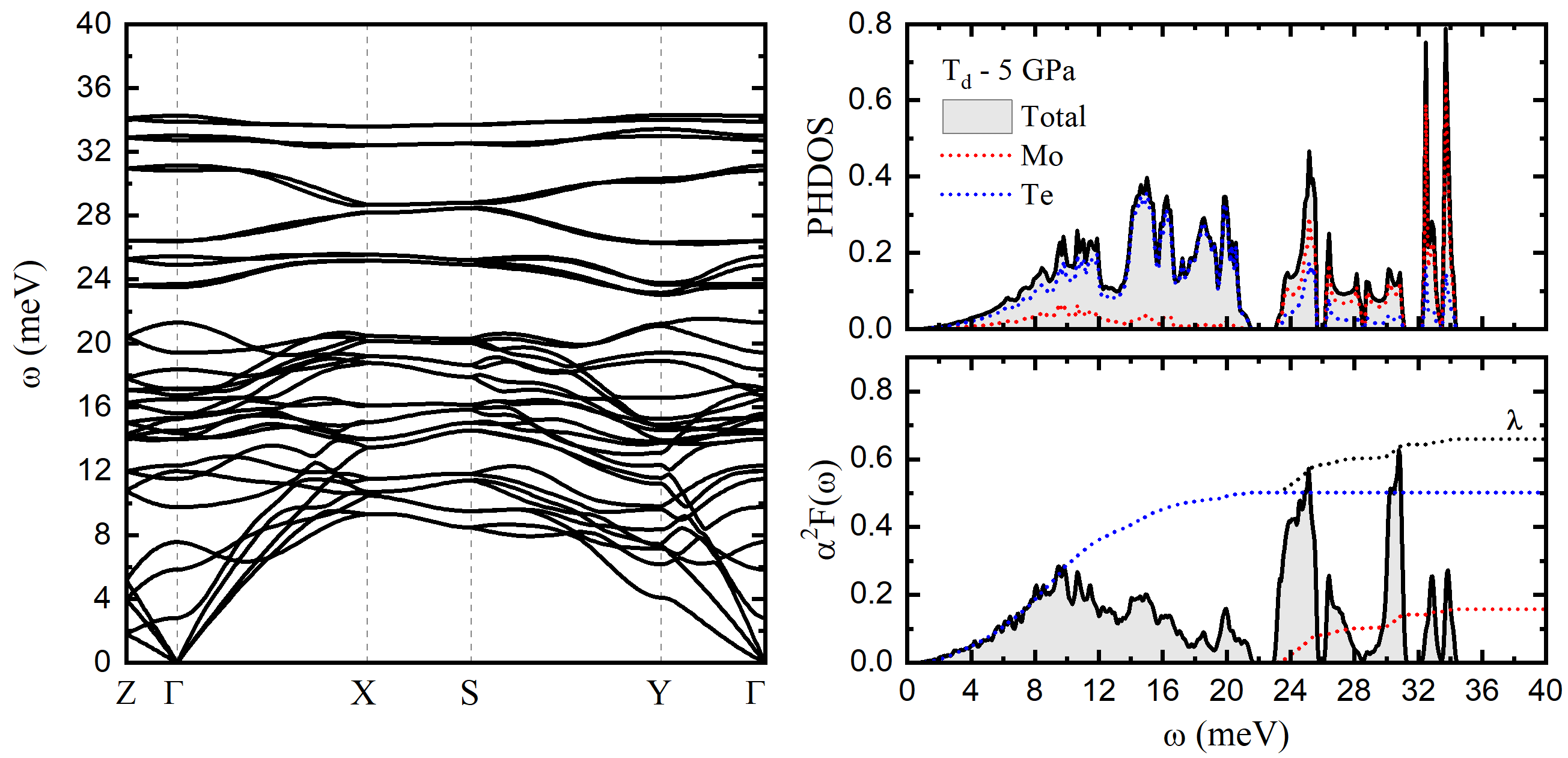}
\includegraphics[width=0.495\linewidth]{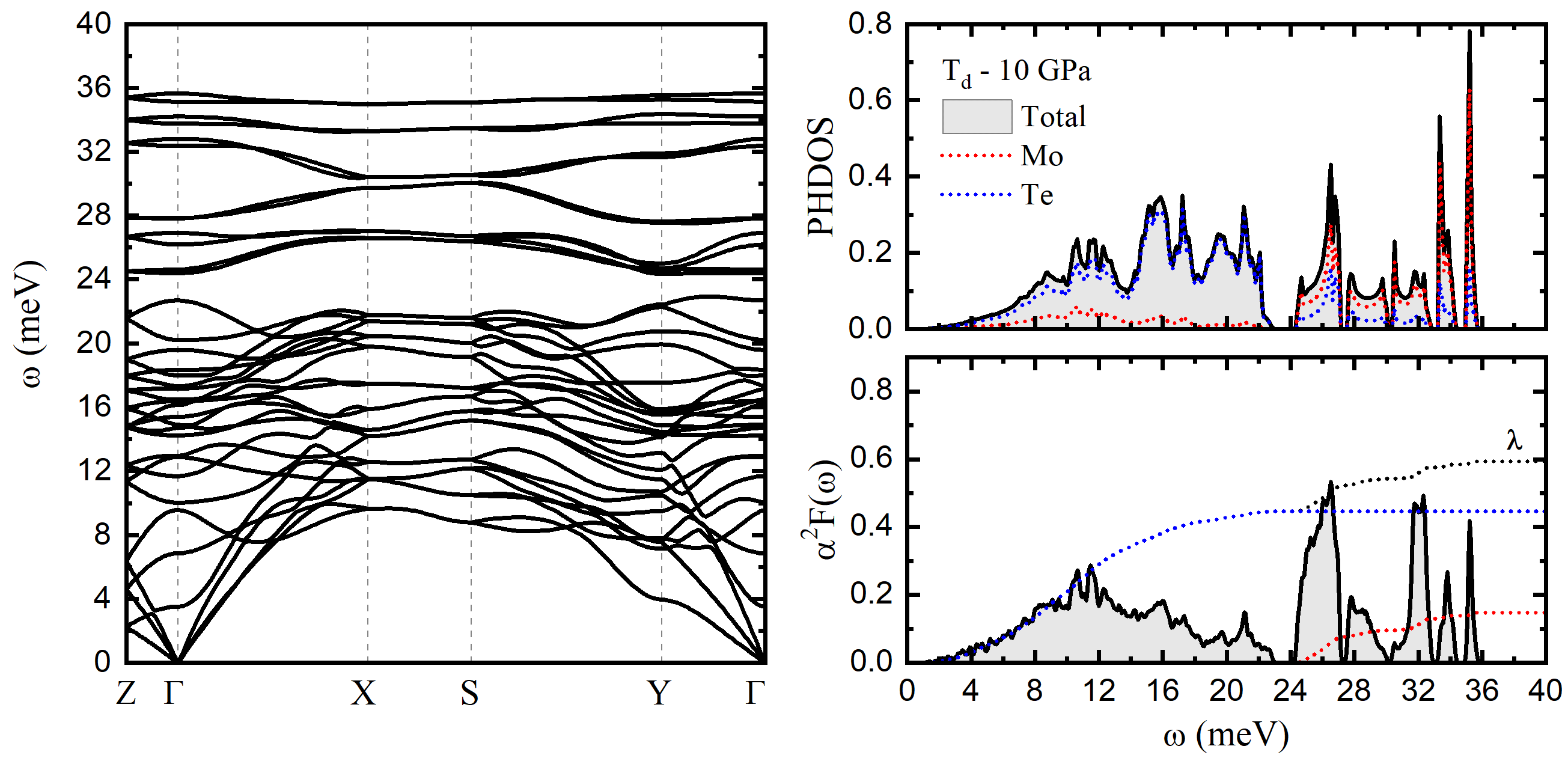}
\caption{\label{fig-S5} Calculated phonon dispersion, PHDOS, isotropic Eliashberg spectral function $\alpha^2F(\omega)$, electron-phonon coupling strength $\lambda(\omega)$ for the T$_d$ phase at various pressures. The total PHDOS (shaded gray area) and the total EPC strength (dotted black line) are decomposed with respect to the vibrations of the Mo (dotted red line) and Te (dotted blue line) atoms. The solid symbols at 0 GPa represent phonon frequencies calculated with SOC.}
\end{figure}	

\begin{figure}[h!]
\centering
\includegraphics[width=0.495\linewidth]{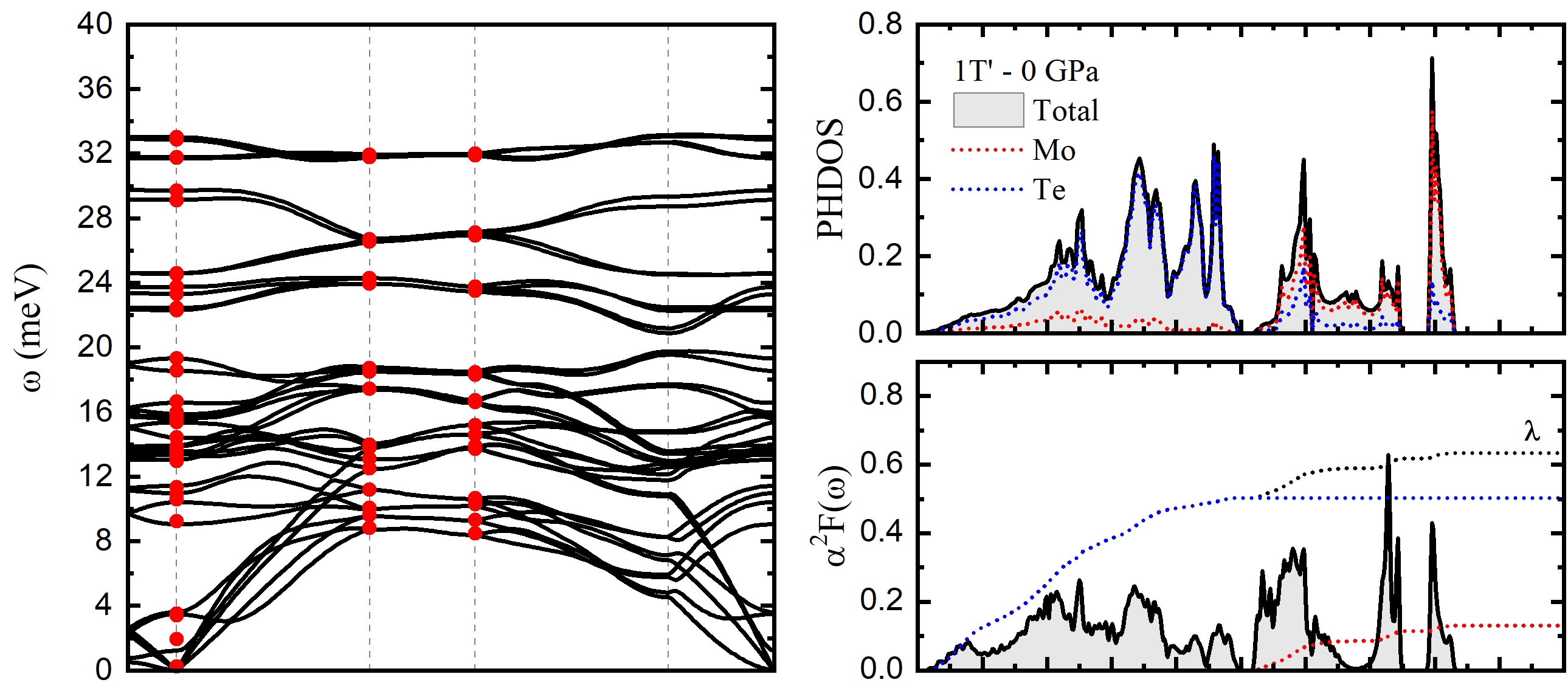}
\includegraphics[width=0.495\linewidth]{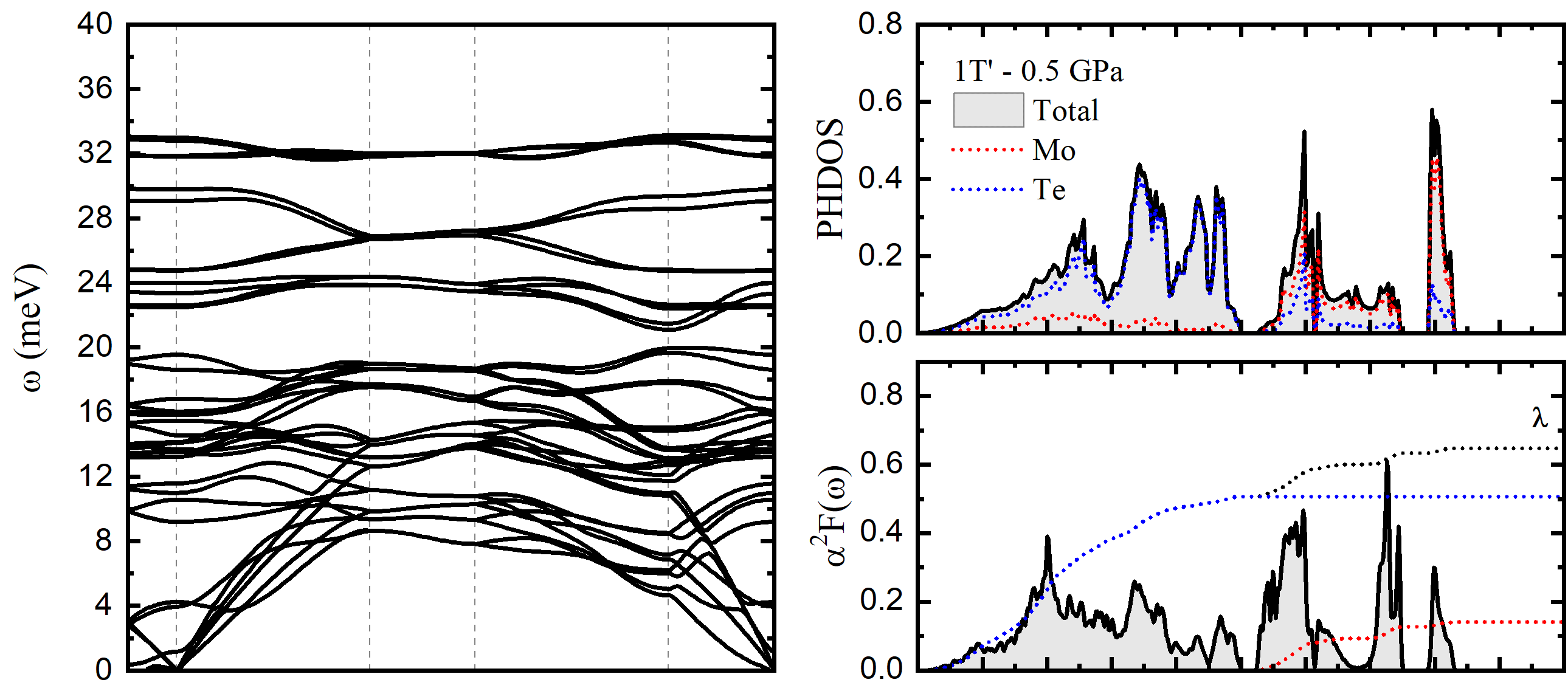}
\includegraphics[width=0.495\linewidth]{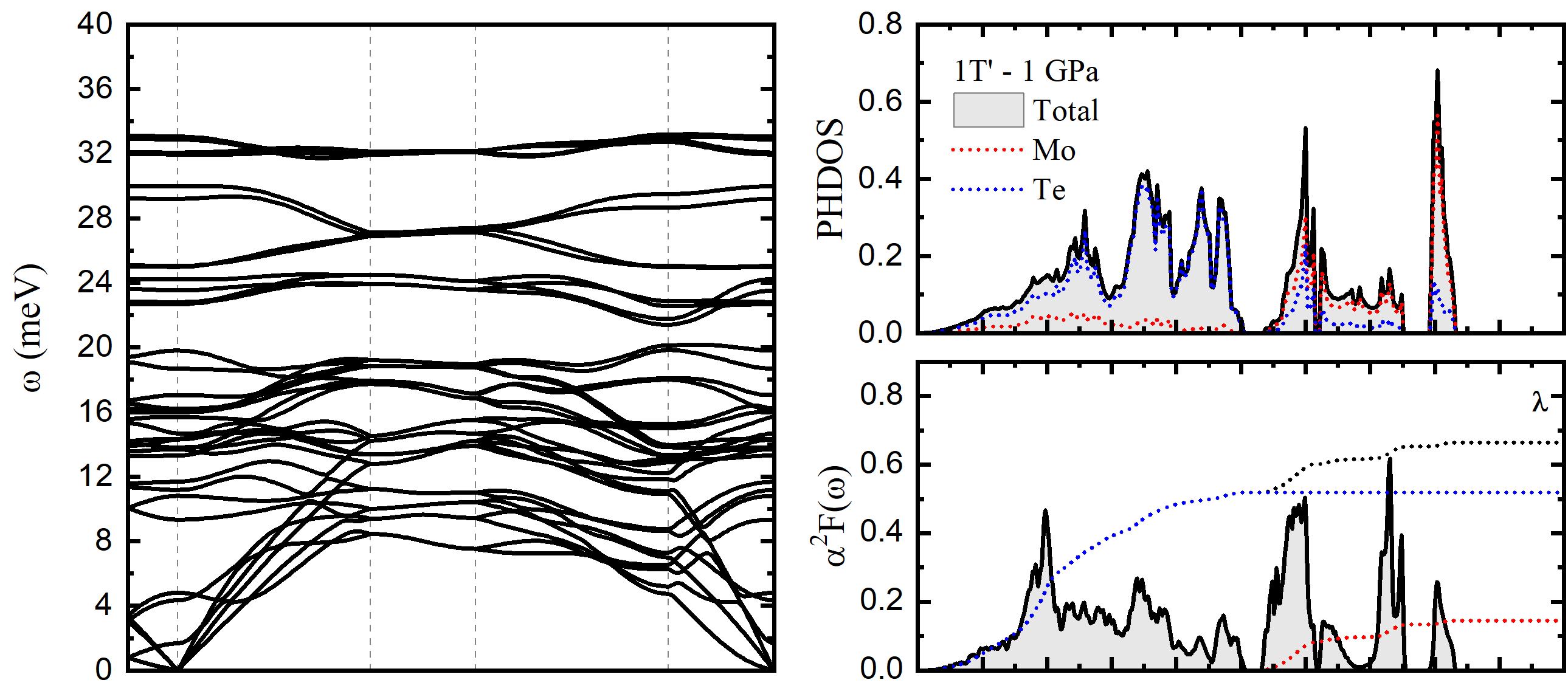}
\includegraphics[width=0.495\linewidth]{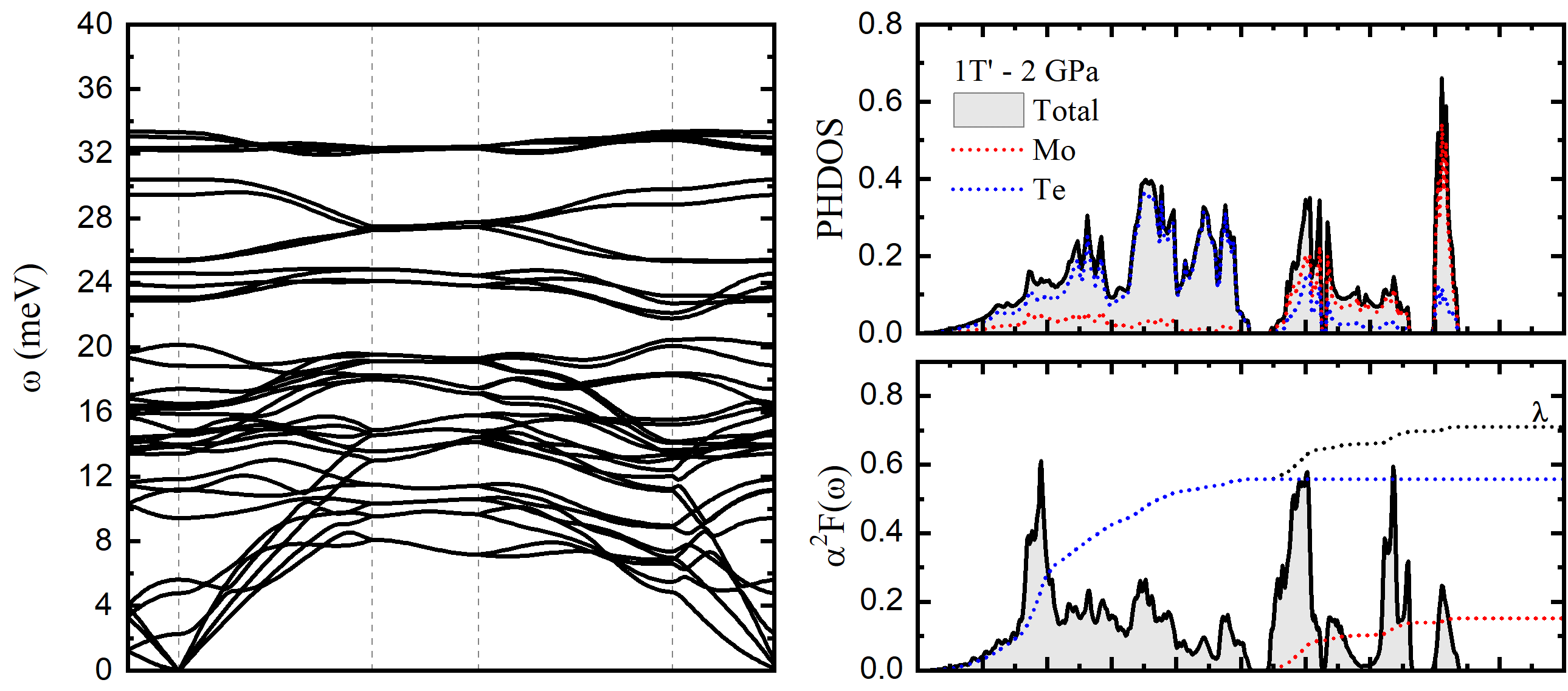}
\includegraphics[width=0.495\linewidth]{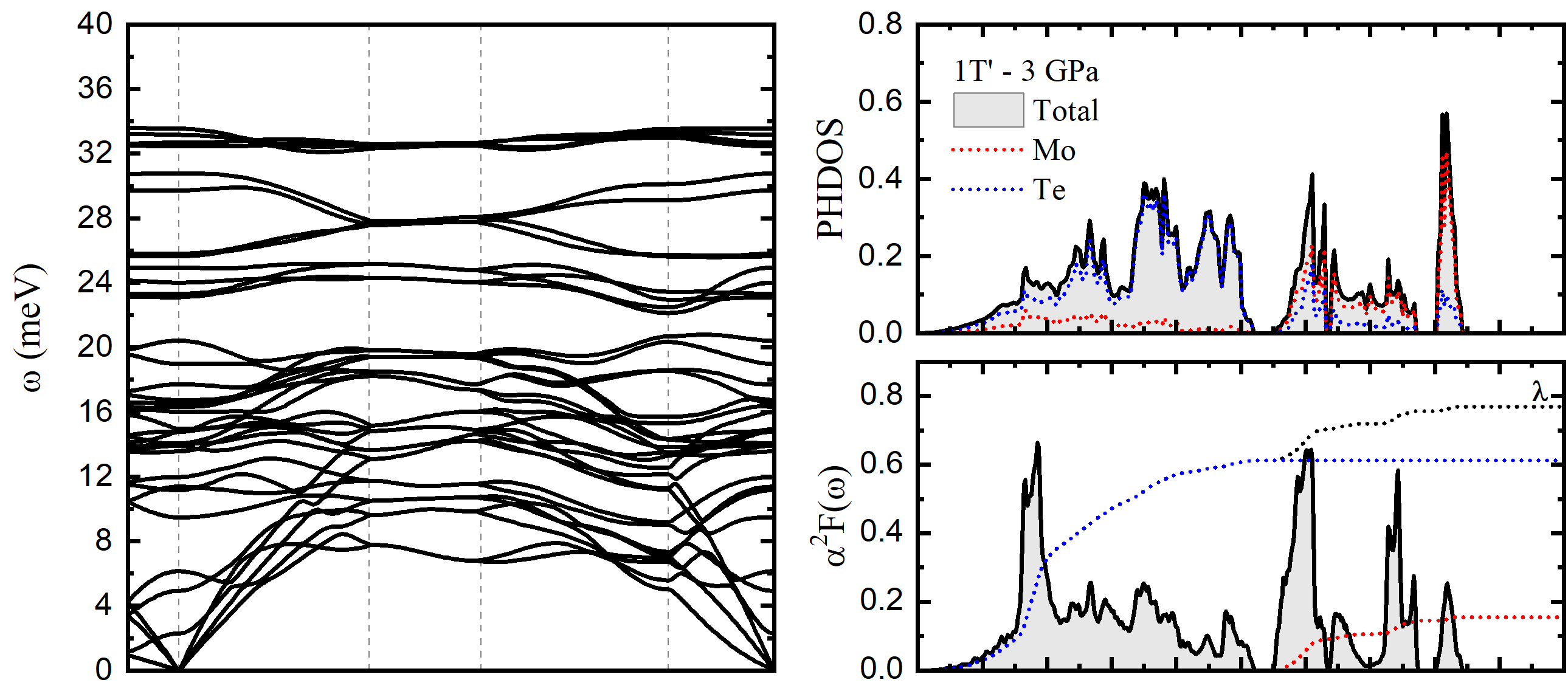}
\includegraphics[width=0.495\linewidth]{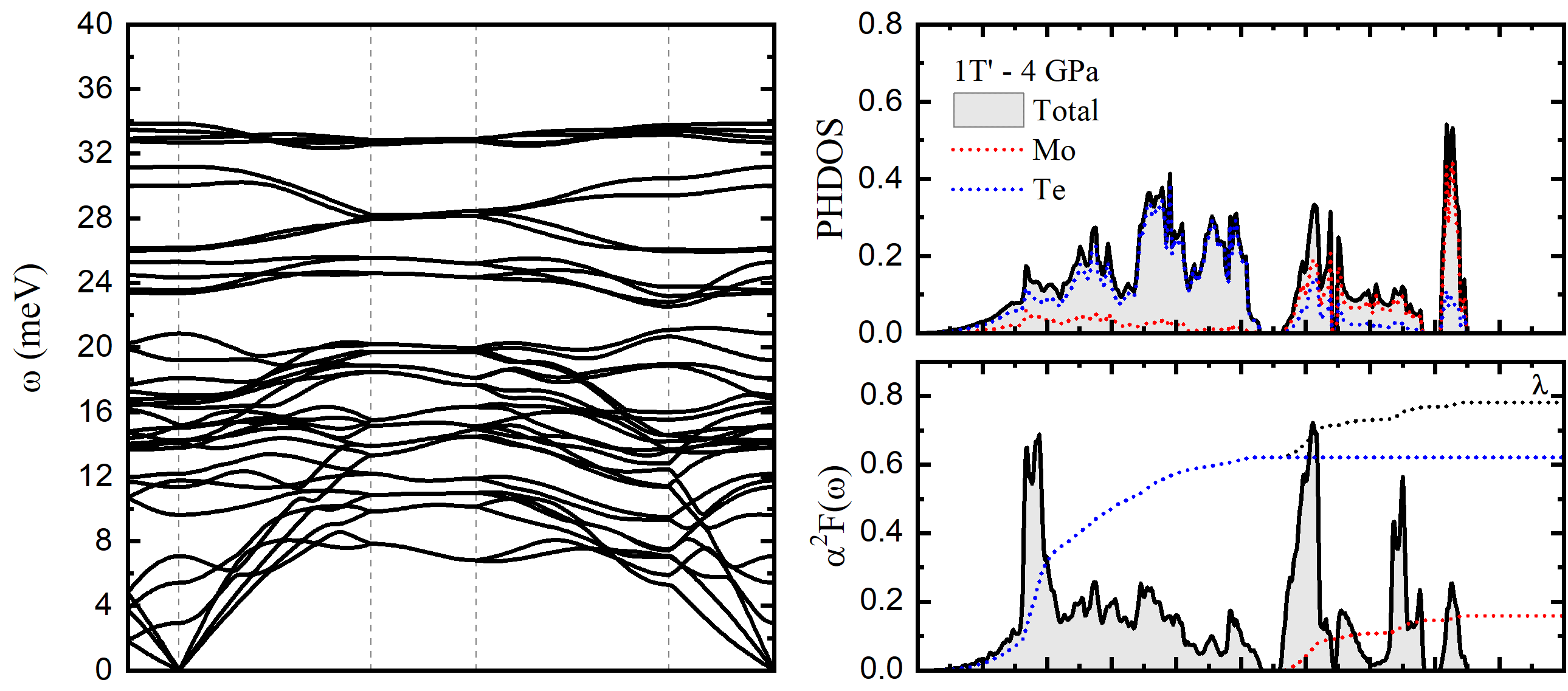}
\includegraphics[width=0.495\linewidth]{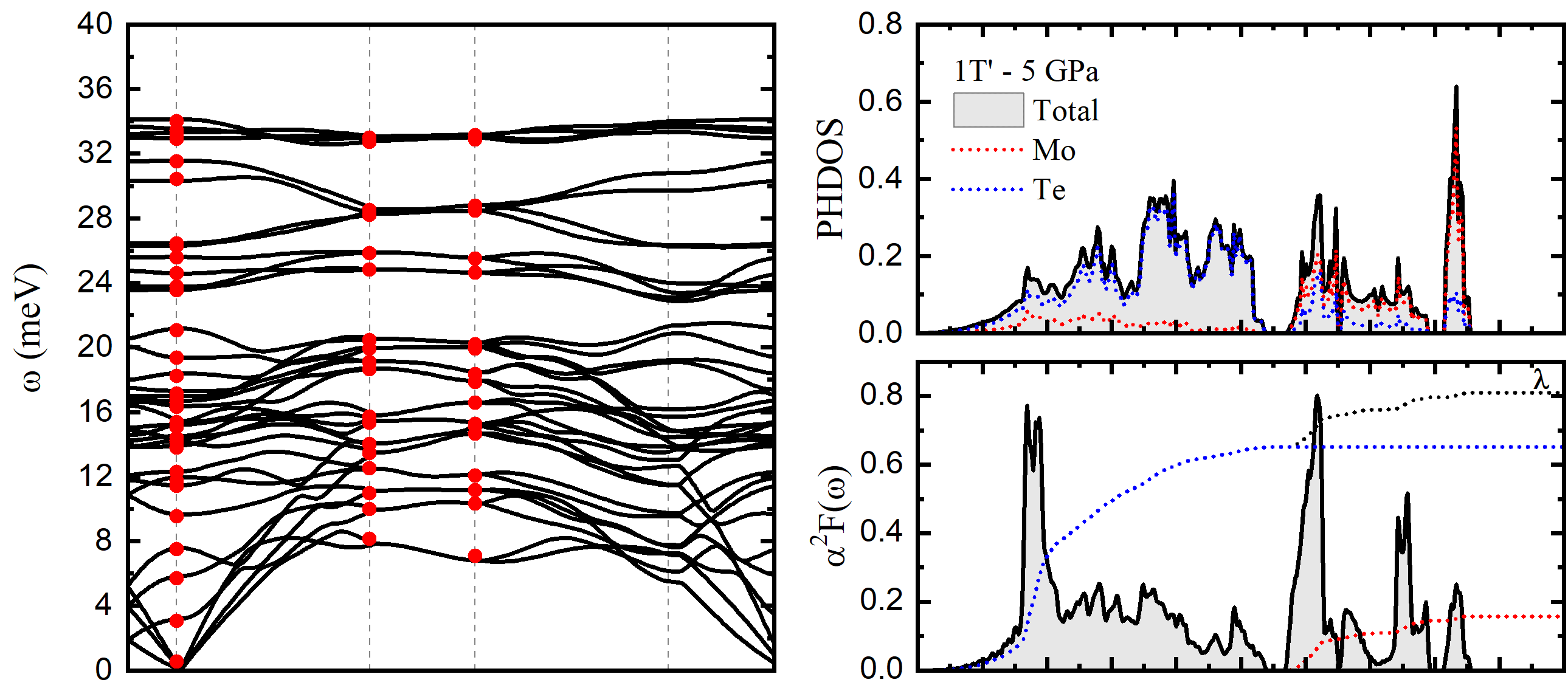}
\includegraphics[width=0.495\linewidth]{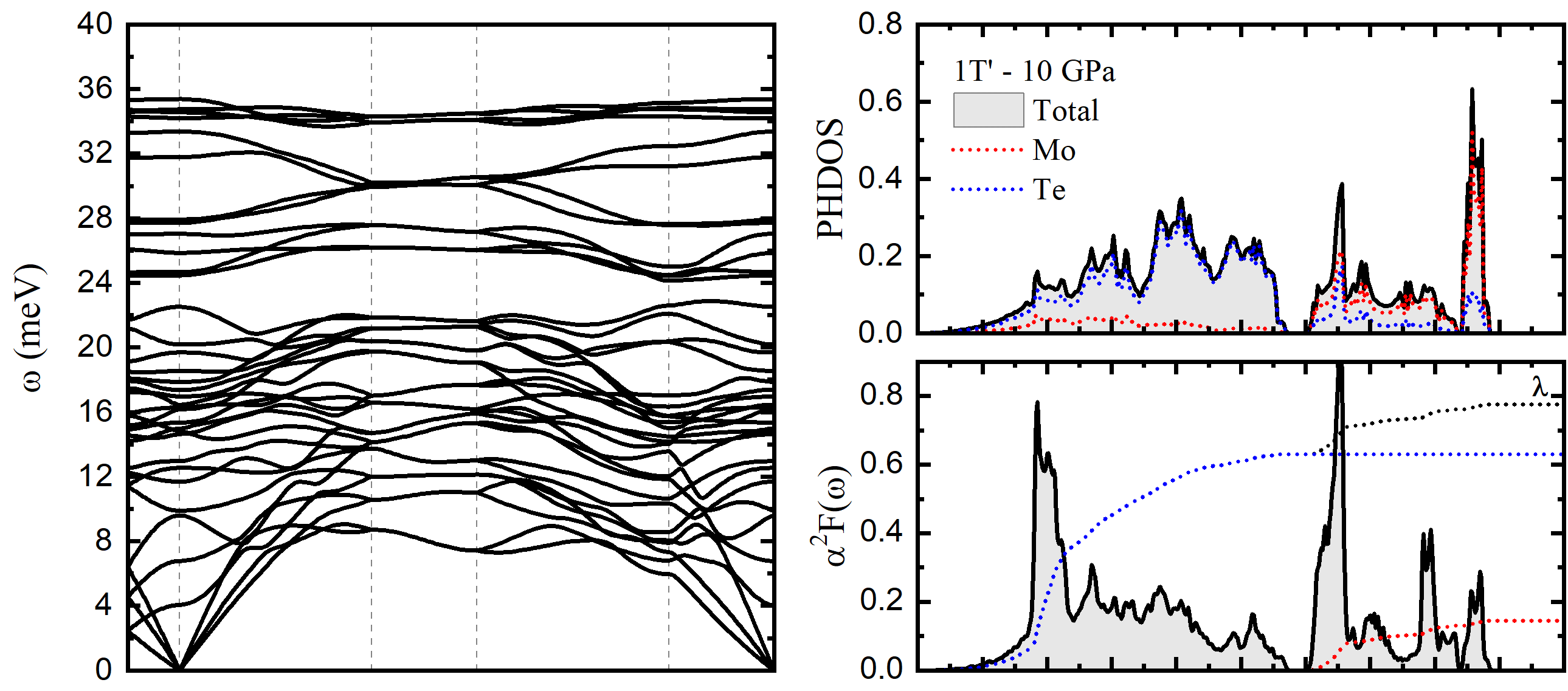}
\includegraphics[width=0.495\linewidth]{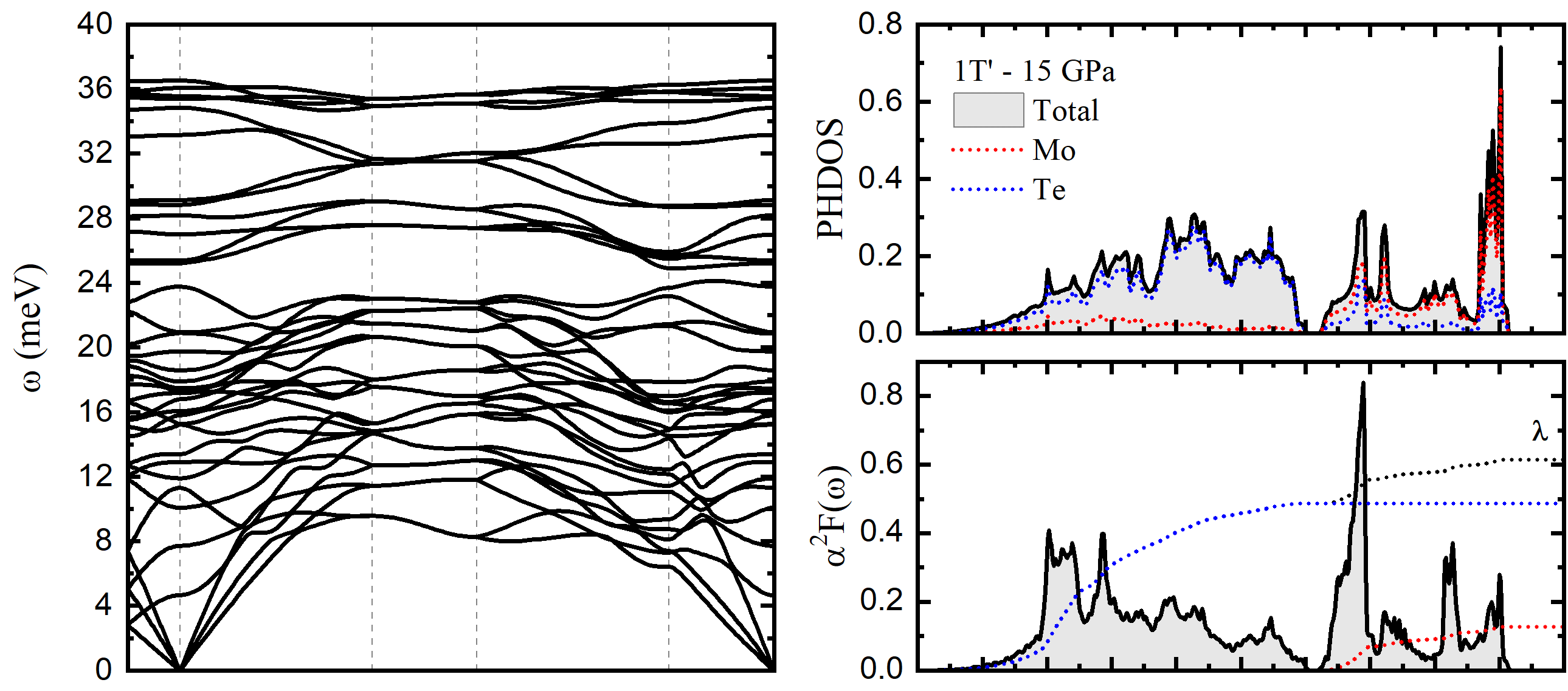}
\includegraphics[width=0.495\linewidth]{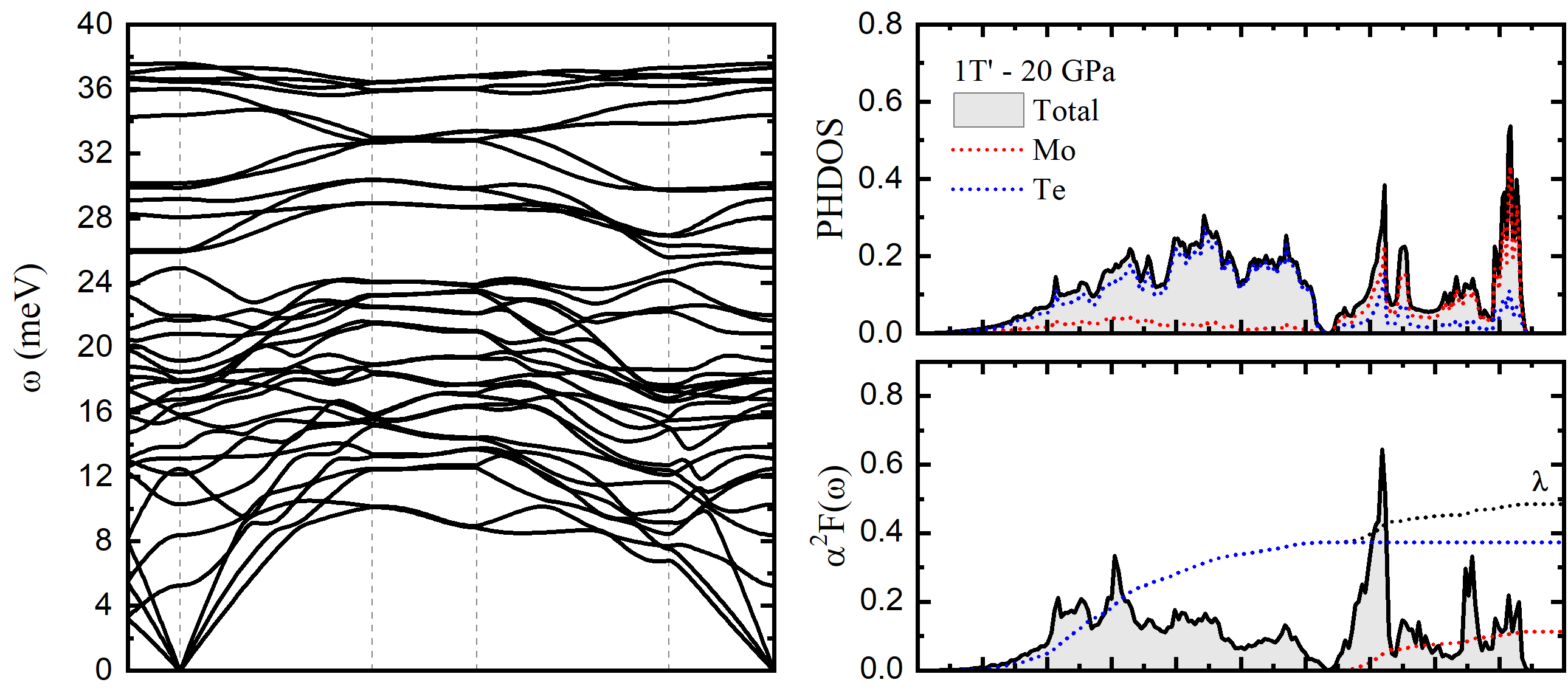}
\includegraphics[width=0.495\linewidth]{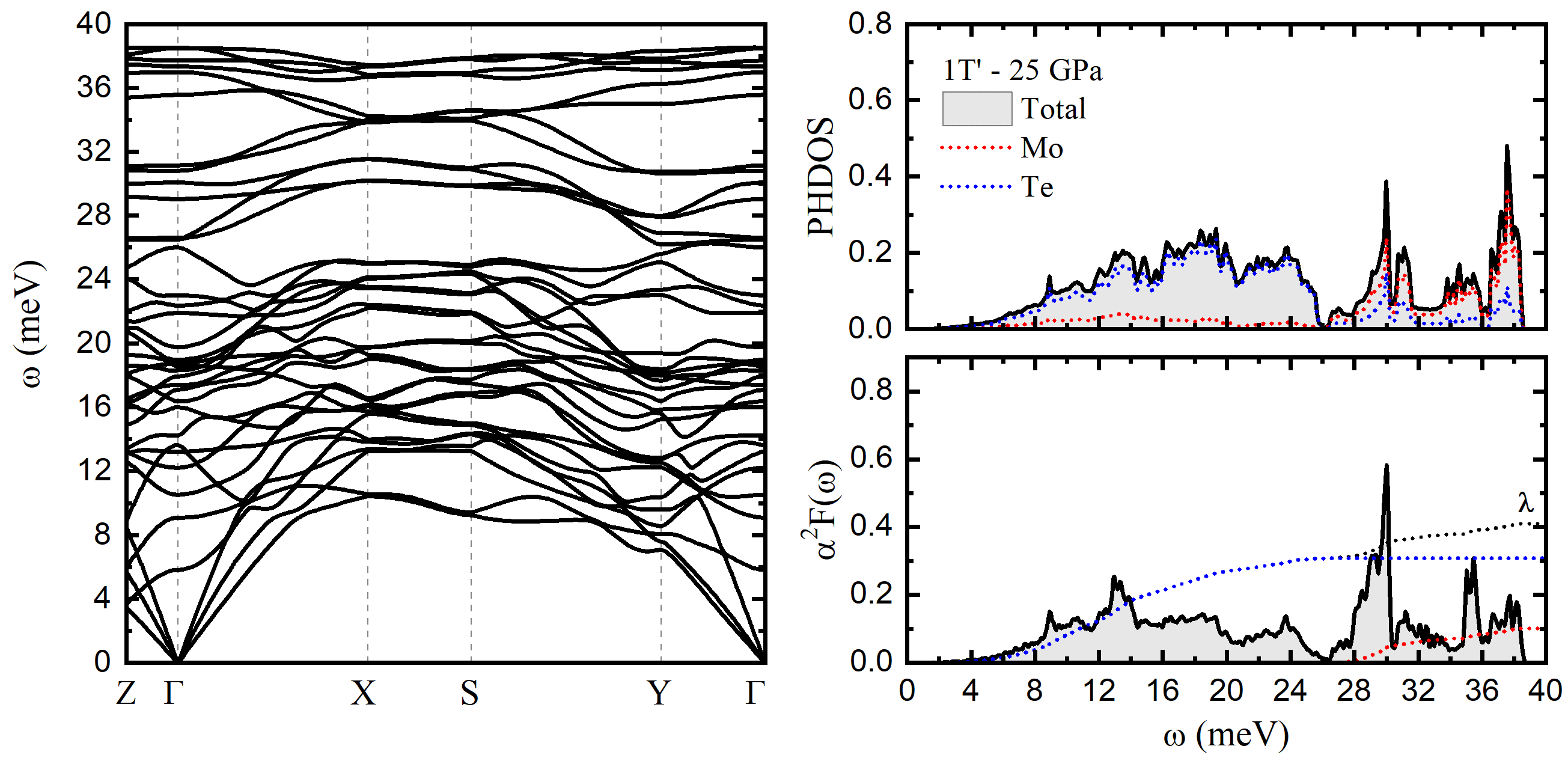}
\includegraphics[width=0.495\linewidth]{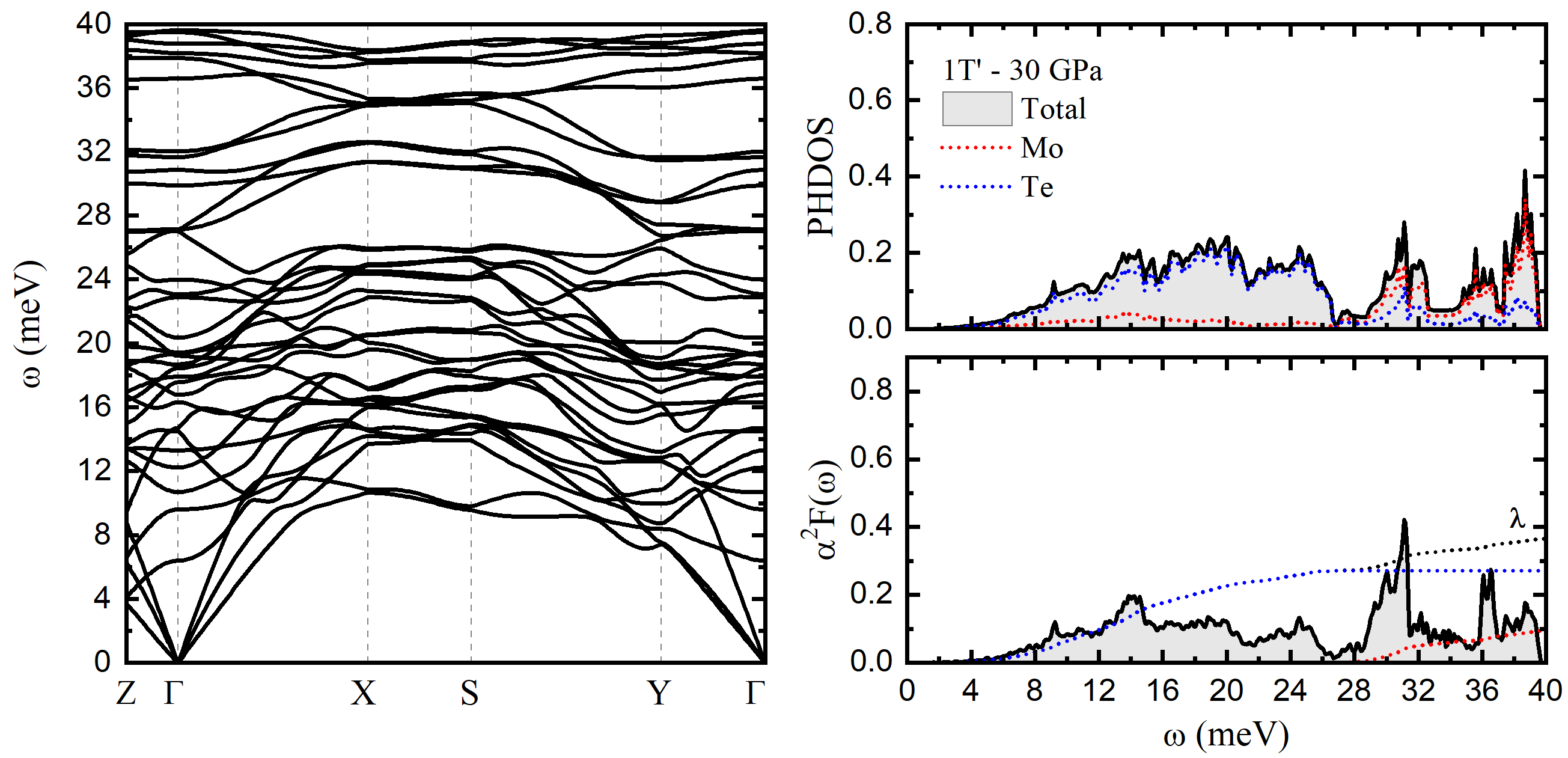}
\caption{\label{fig-S6} Calculated phonon dispersion, PHDOS, isotropic Eliashberg spectral function $\alpha^2F(\omega)$, electron-phonon coupling strength $\lambda(\omega)$ for the 1T$^\prime$ phase at various pressures. The total PHDOS (shaded gray area) and the total EPC strength (dotted black line) are decomposed with respect to the vibrations of the Mo (dotted red line) and Te (dotted blue line) atoms. The solid red symbols at 0 GPa represent phonon frequencies calculated with SOC.}
\end{figure}

\begin{figure}[h!]
	\centering
	\includegraphics[width=0.99\columnwidth]{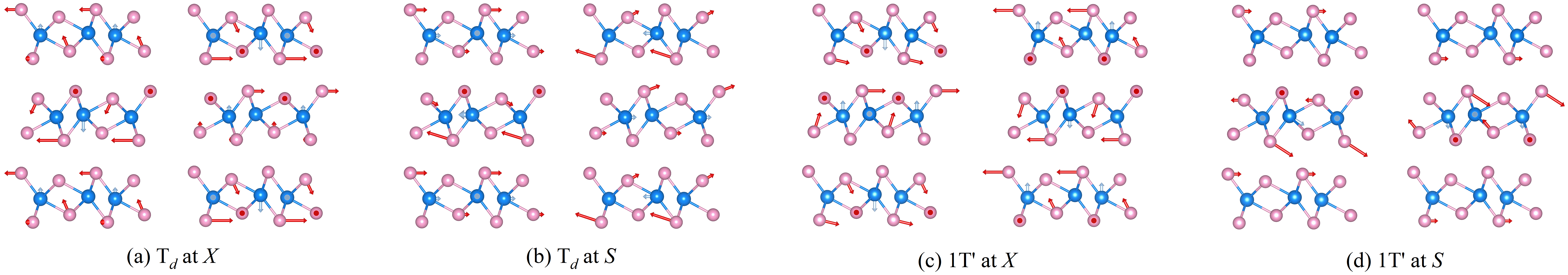}
	\caption{\label{fig-S7} (a-d) Atomic displacement patterns for the lowest energy degenerate mode at the $X$ and $S$ high-symmetry points in the BZ for the T$_d$ and 1T$^\prime$ phases at 0 GPa.}
\end{figure}

\begin{figure}[h!]
	\centering
	\includegraphics[width=0.99\columnwidth]{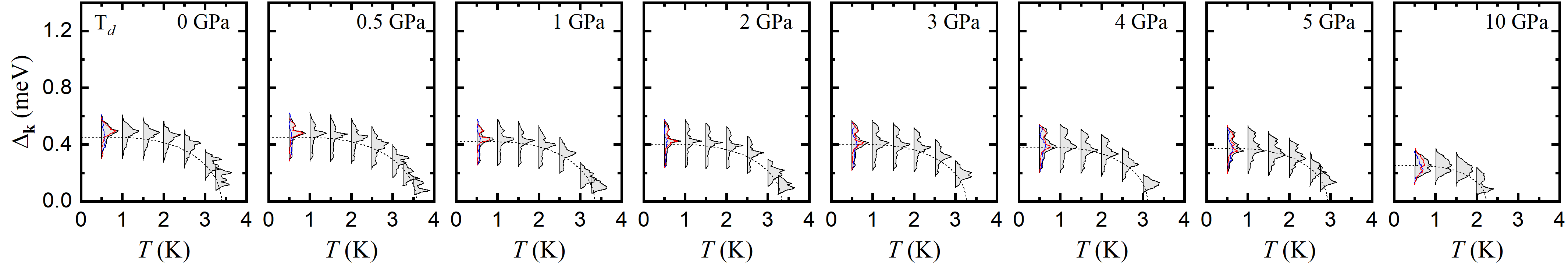}
	\caption{\label{fig-S8} Energy distribution of the anisotropic superconducting gap $\Delta_\textbf{k}$ as a function of temperature for the T$_d$ phase at various pressures. The dashed lines are fits obtained by solving numerically the BCS gap equation using the average $\Delta_0$ and $T_{\rm c}$ from our first-principles calculations.}
\end{figure}

\begin{figure}[!]
	\centering
	\includegraphics[width=0.9\columnwidth]{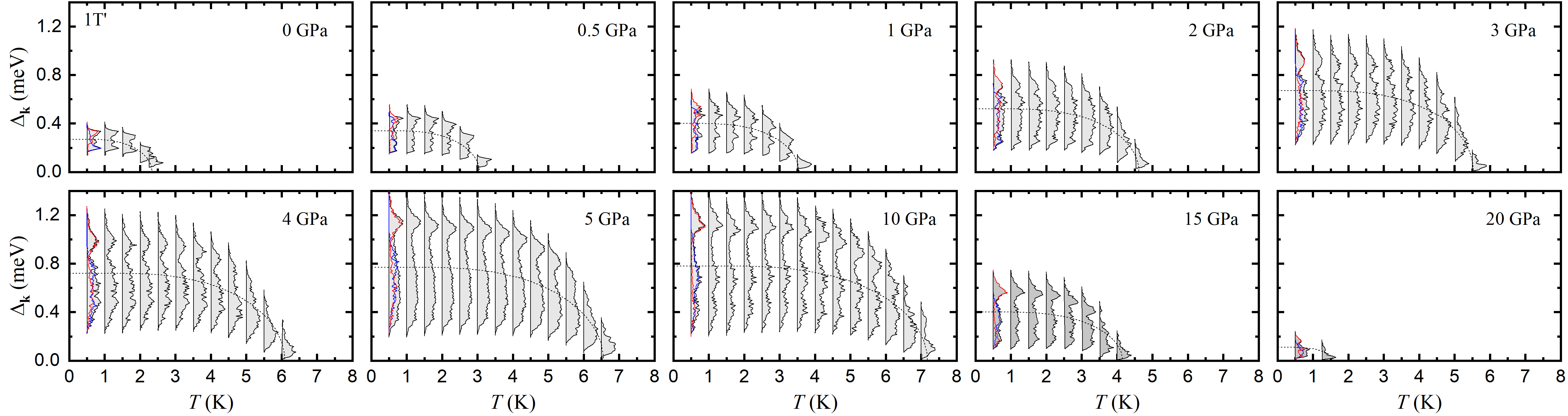}
	\caption{\label{fig-S9} Energy distribution of the anisotropic superconducting gap $\Delta_\textbf{k}$ as a function for the 1T$^\prime$ phase of temperature at various pressures. The dashed lines are fits obtained by solving numerically the BCS gap equation using the average $\Delta_0$ and $T_{\rm c}$ from our first-principles calculations.}
\end{figure}


\begin{figure}[h!]
	\centering
	\includegraphics[width=0.65\columnwidth]{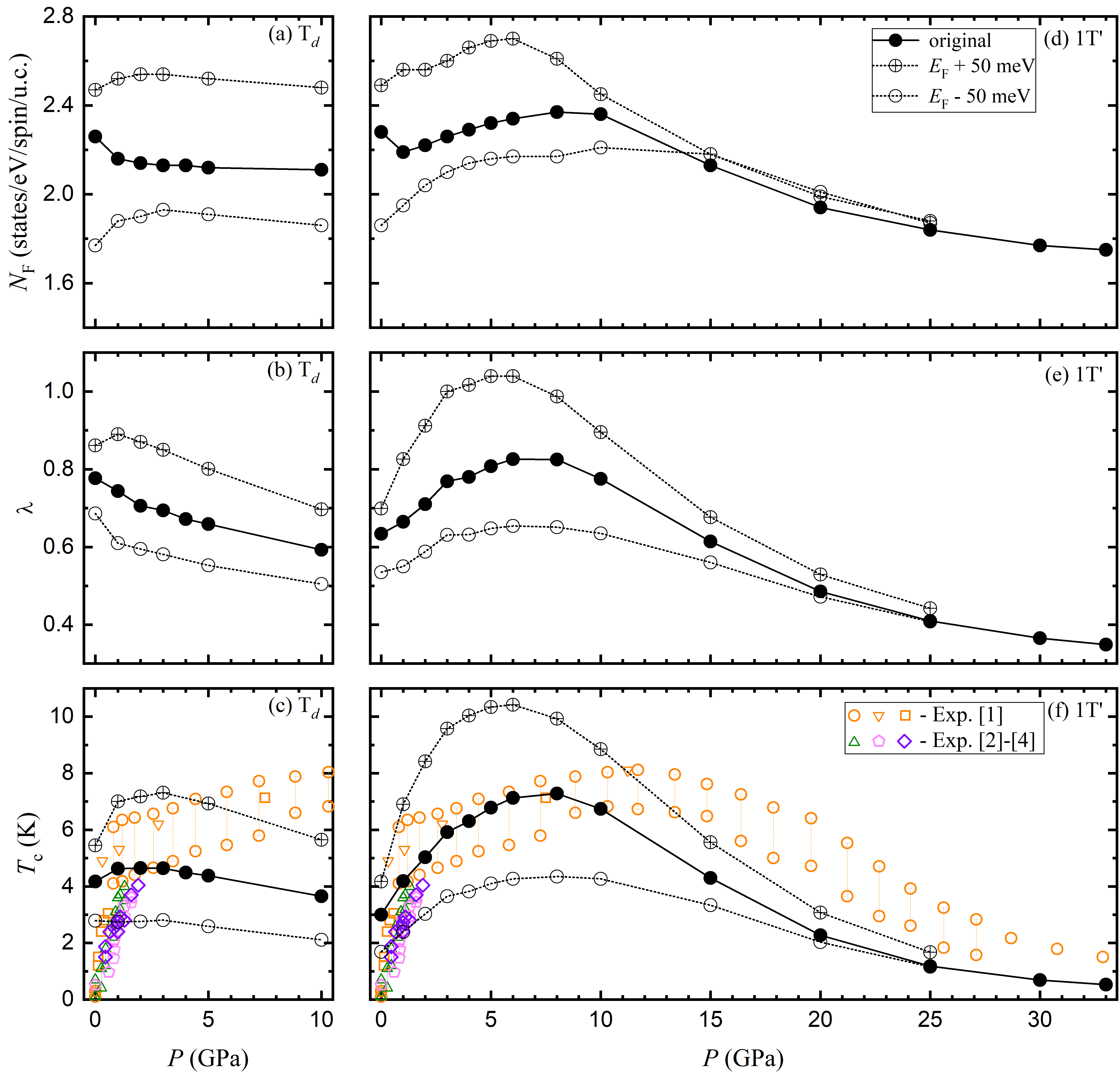}
	\caption{\label{fig-S10} Sensitivity of the calculated superconducting properties of the T$_d$ and 1T$^\prime$ phases to variations in the density of states at the Fermi level. (a, d) $N_{\rm F}$, (b, e) $\lambda$, and (c, f) $T_{\rm c}$ as a function of pressure for the T$_d$ and 1T$^\prime$ phases. The $T_{\rm c}$ is calculated using the isotropic ME theory. Open and crossed black circles correspond to a -50~meV and +50~meV rigid shift of the Fermi level with respect to the original data (filled black circles). The change in the $N_{\rm F}$ has a significant impact on the calculated $T_{\rm c}$, mostly due to its effect on $\lambda$. These results constitute theoretical lower and upper bound for $\lambda$ and $T_{\rm c}$, and account for some of the differences in the band structure between first-principles calculations and angular-resolved photoemission spectroscopy experiments. The experimental $T_{\rm c}$ values are from Ref.~[\onlinecite{QI}] (orange circles and triangles from electrical resistivity  and orange squares from magnetization measurements),  Ref.~[\onlinecite{TAKAHASHI}] (green triangles from electrical resistivity measurements),  Ref~[\onlinecite{HEIKES}] (magenta pentagons from neutron scattering measurements), and Ref.~[\onlinecite{GUGUCHIA}] (purple diamonds from AC-susceptibility measurements).}
\end{figure}

\begin{figure}[h!]
	\centering
	\includegraphics[width=0.99\columnwidth]{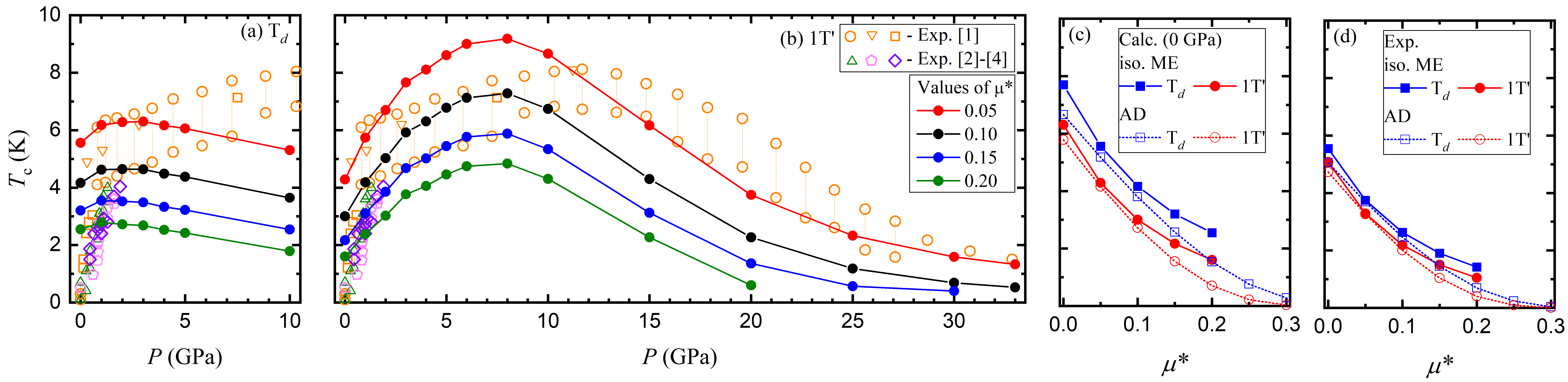}
	\caption{\label{fig-S121} Variation of the superconducting $T_{\rm c}$ as function of pressure calculated using the isotropic ME theory for different values of the Coulomb pseudopotential $\mu^*$ in the (a) T$_d$ and (b) 1T$^\prime$ phases. Comparison between the superconducting $T_{\rm c}$ calculated using the isotropic ME theory (solid symbols) and from the Allen-Dynes formula (open symbols) at the (c) calculated lattice constants at 0 GPa and (d) experimental lattice constants at ambient pressure for various $\mu^*$ values. }
\end{figure}

\begin{table}[h!]
	\resizebox{0.68\columnwidth}{!}{%
		\begin{tabular}{|c|c|c|c|c|c||c|c|c|c|c|c|c|c|}
			\hline
			\multicolumn{6}{|c||}{T$_d$ ($\Gamma$ = 12$A_1$ + 6$A_2$ + 6$B_1$ + 12$B_2$)} & \multicolumn{8}{|c|}{1T$^\prime$ ($\Gamma$ = 12$A_g$ + 6$A_u$ + 6$B_g$ + 12$B_u$)}\\
			\hline
			Rep.  &  \multicolumn{2}{|c|}{Present}  & \multicolumn{2}{|c|}{Ref.\cite{ZHANG}} & {Ref.\cite{CHEN2}}  &  Rep.  & \multicolumn{2}{|c|} {Present}  & \multicolumn{2}{|c|}{Ref.\cite{ZHANG}} & {Ref.\cite{CHEN2}} & \multicolumn{2}{|c|}{Ref.\cite{SONG}} \\
			&   \multicolumn{2}{|c|} {work}  & \multicolumn{2}{|c|}{} &   &   & \multicolumn{2}{|c|} {work}  & \multicolumn{2}{|c|}{} & & \multicolumn{2}{|c|}{} \\
			\hline
			&Calc.    &Exp.    & Calc.& Exp. & Calc. &   & Calc.    & Exp.   & Calc. & Exp. & Calc. & Exp. & Calc. \\
			&Lattice  &Lattice  &      &       &       &   & Lattice  & Lattice &       &       &       &       &       \\
			\hline
			$A_1$   & 0.0   & 0.0   & --   & 0        & 0.0   & $A_g$ & 74.5  & 75.0  & 77    & 78       & 73.9     & 78    & 78.4   \\
			(IR+R)  & 11.1  & 15.5  & 13   & 14       & 9.2   & (R)   & 85.4  & 87.9  & 88    & 89       & 86.6     & 90    & 90.7   \\
			& 73.2  & 75.2  & 77   & 78       & 73.5  &       & 109.1 & 109.6 & 111   & 114      & 109.0    & 112   & 114.4  \\
			& 109.3 & 109.7 & 113  & 115      & 110.0 &       & 111.7 & 114.9 & 116   & 119      & 113.0    &       & 118.3  \\
			& 123.6 & 126.8 & 128  & 129      & 123.8 &       & 126.0 & 128.1 & 128   & 133      & 125.6    & 129   & 132.2  \\
			& 127.2 & 129.6 & 132  & 133      & 127.9 &       & 128.0 & 130.3 & 128   & 134      & 128.0    &       & 134.0  \\
			& 133.7 & 135.7 &      & 142      & 138.5 &       & 149.7 & 152.6 & 158   & 157      & 149.5    &       & 156.0  \\
			& 156.0 & 159.5 & 165  & 165      & 157.7 &       & 155.9 & 159.5 & 164   & 166      & 154.7    & 163   & 164.1  \\
			& 198.3 & 203.2 &      & 211      & 201.9 &       & 234.7 & 243.2 &       & 240      & 227.2    &       & 244.4  \\
			& 236.8 & 245.4 &      & 248      & 232.3 &       & 239.5 & 249.8 &       & 252      & 235.0    &       & 250.4  \\
			& 256.2 & 261.4 &      & 267      & 262.0 &       & 255.8 & 262.0 &       & 268      & 258.0    & 261   & 268.6  \\
			& 266.3 & 269.1 &      & 276      & 269.2 &       & 256.2 & 262.3 &       & 271      & 260.6    &       & 269.9  \\
			\hline
			$A_2$   & 27.9  & 29.7  &     & 36        & 32.4  & $A_u$ & 0.0   & 0.0   &       & 0        & 0.0      &       & 11.8   \\
			(R)     & 91.8  & 93.8  & 96  & 98        & 94.1  & (IR)  & 27.3  & 29.7  &       & 36       & 29.3     &       & 37.8   \\
			& 105.7 & 105.2 & 108 & 112       & 107.4 &       & 108.9 & 109.9 &       & 115      & 110.3    &       & 114.4  \\
			& 109.8 & 110.1 & 112 & 115       & 111.9 &       & 111.7 & 114.2 &       & 119      & 113.8    &       & 118.3  \\
			& 180.1 & 181.0 & 188 & 192       & 184.3 &       & 179.8 & 181.0 &       & 192      & 181.9    &       & 189.2  \\
			& 189.2 & 194.8 & 194 & 200       & 189.7 &       & 180.8 & 181.9 &       & 194      & 183.5    &       & 190.3  \\
			\hline
			$B_1$   & 0.0   & 0.0   &     & 0         & 0.0   & $B_u$ & 0.0   & 0.0   &       & 0        & 0.0      &       & -4.9   \\
			(IR+R)  & 28.3  & 32.9  &     & 37        & 27.2  & (IR)  & 0.0   & 0.0   &       & 0        & 0.0      &       & -2.3   \\
			& 83.9  & 87.4  & 88  & 89        & 87.7  &       & 15.5  & 12.5  &       & 11       & 15.3     &       & 15.8   \\
			& 112.0 & 114.1 & 118 & 119       & 112.7 &       & 28.4  & 31.9  &       & 37       & 31.45    &       & 37.7   \\
			& 116.5 & 120.1 & 118 & 121       & 115.0 &       & 116.2 & 120.2 &       & 120      & 117.8    &       & 121.0  \\
			& 127.4 & 128.9 & 129 & 134       & 128.3 &       & 132.9 & 127.5 &       & 129      & 126.3    &       & 129.4  \\
			& 128.4 & 130.3 & 131 & 136       & 130.0 &       & 128.8 & 130.8 &       & 136      & 131.8    &       & 134.5  \\
			& 150.0 & 152.7 & 159 & 159       & 147.4 &       & 134.2 & 136.8 &       & 142      & 138.0    &       & 141.0  \\
			& 198.0 & 203.0 &     & 211       & 201.0 &       & 197.8 & 203.5 &       & 211      & 200.8    &       & 207.2  \\
			& 238.0 & 247.5 &     & 248       & 234.0 &       & 198.3 & 203.9 &       & 212      & 202.1    &       & 207.8  \\
			& 255.8 & 261.1 &     & 270       & 257.1 &       & 265.1 & 268.0 &       & 275      & 268.6    &       & 275.2  \\
			& 256.3 & 269.0 &     & 277       & 262.8 &       & 266.3 & 268.9 &       & 276      & 270.3    &       & 276.8  \\
			\hline
			$B_2$   & 0.0   & 0.0   &     & 0         & 0     & $B_g$ & 87.8  & 90.3  & 90    & 93       & 89.6     &       & 90.7   \\
			(IR+R)  & 88.8  & 90.1  & 92  & 93        & 88.3  &  (R)  & 91.6  & 94.1  & 94    & 98       & 93.5     & 95    & 95.8   \\
			& 107.3 & 107.7 & 111 & 115       & 109.8 &       & 104.6 & 105.1 & 107   & 112      & 106.6    & 109   & 110.6  \\
			& 112.4 & 114.4 & 115 & 119       & 115.5 &       & 107.1 & 107.6 & 111   & 115      & 108.5    &       & 113.3  \\
			& 180.7 & 181.4 &     & 194       & 185.0 &       & 187.8 & 193.4 & 191   & 200      & 190.4    & 195   & 198.4  \\
			& 190.7 & 169.7 &     & 205       & 194.7 &       & 191.5 & 198.1 & 193   & 204      & 196.6    &       & 204.4  \\
			\hline
		\end{tabular}
	}
		\caption{\label{Table1}
		Comparison between calculated and experimental phonon frequencies (in cm$^{-1}$) for the T$_d$ and 1T$^\prime$ phases at $\Gamma$-point at 0~GPa.}
\end{table}
	
\FloatBarrier